\documentclass[a4paper,11pt]{article}
\usepackage{jheppub}
\usepackage[utf8]{inputenc}
\usepackage{{amsmath,amsfonts,latexsym, amstext, amssymb, amsthm}}
\usepackage{xspace} 
\usepackage{slashed}
\usepackage[bf]{caption}
\usepackage{subcaption}
\usepackage{comment}
 \usepackage{hyperref}
\usepackage{dsfont}
\usepackage{placeins}

\definecolor{labelcolor}{rgb}{0,0,0}

\usepackage[textsize=footnotesize,textwidth=2.25cm]{todonotes}

\usepackage{feynmp}
\usepackage{ifpdf}

\ifpdf
\else
\fi

\usepackage{xkeyval}
\input{picturemacro}

\usepackage{tikz}
\usetikzlibrary{shapes}

\newlength{\vacuumradius}
\newlength{\onshellradius}
\tikzstyle{db}=[circle, black, fill=black, minimum width=\onshellradius, draw, inner sep=0pt]
\tikzstyle{dw}=[circle, black, fill=white, minimum width=\onshellradius, draw, inner sep=0pt]
\tikzstyle{dvac}=[circle, black, fill=lightgray, minimum width=\vacuumradius, inner sep=0pt]

\def\blobhdist{0.4}
\def\blobvdist{0.1}
\def\extradist{0.12}
\def\blobheight{0.7}
\def\blobwidth{\blobhdist*1.5}

\tikzstyle{twoblob}=[ellipse, black, fill=white, minimum width=\blobwidth cm, minimum height=\blobheight cm, draw, inner sep=0pt]
\tikzstyle{threeblob}=[ellipse, black, fill=white, minimum width=\blobhdist cm + \blobwidth cm, minimum height=\blobheight cm, draw, inner sep=0pt]

\newcommand{\bloblabelsize}{\scriptsize}

\newcommand{\drawvline}[2]{
        \draw (#1*\blobhdist-1*\blobhdist,0) -- (#1*\blobhdist-1*\blobhdist,#2*\blobvdist +#2*\blobheight +\blobvdist);
        \draw[white] (#1*\blobhdist-1*\blobhdist-\extradist,0) -- (#1*\blobhdist-1*\blobhdist +\extradist,0);}
\newcommand{\drawtwoblob}[3]{
\node[twoblob] at (#1*\blobhdist-1*\blobhdist+0.5*\blobhdist,#2*\blobvdist +#2*\blobheight-0.5*\blobheight) {\bloblabelsize #3};
}        
\newcommand{\drawthreeblob}[3]{
\node[threeblob] at (#1*\blobhdist-1*\blobhdist +1*\blobhdist,#2*\blobvdist +#2*\blobheight-0.5*\blobheight) {\bloblabelsize #3};
}

\newcommand{\la}{\langle}
\newcommand{\ra}{\rangle}
\newcommand{\ab}[1]{\la #1 \ra}
\renewcommand{\sb}[1]{[ #1 ]}

\newcommand{\lambdat}{\tilde{\lambda}}

\newcommand{\ZZ}{\ensuremath{\mathbb{Z}}}
\renewcommand{\SS}{\ensuremath{\mathbb{S}}}

\newcommand{\PP}{\ensuremath{\mathbb{P}}}

\newcommand{\SU}[1]{\ensuremath{SU(#1)}\xspace}
\newcommand{\SO}[1]{\ensuremath{SO(#1)}\xspace}

\DeclareMathOperator{\idm}{\mathds{1}}

\newcommand{\YM}{{\mathrm{\scriptscriptstyle YM}}}

\DeclareMathOperator{\tr}{tr}
\newcommand{\Tr}{\tr}

\DeclareMathOperator{\phaneq}{\phantom{{}=}}

\newcommand{\e}{\operatorname{e}}

\newcommand{\ket}[1]{\mathopen{\mid}{}#1{}\mathclose{\rangle}}

\newcommand{\bra}[1]{\mathopen{\langle}{}#1{}\mathclose{\mid}}

\newcommand{\eqndot}{\, .}
\newcommand{\eqncom}{\, ,}

\newcommand{\de}{\operatorname{d}\!}

\newcommand{\tabref}[1]{table~\ref{#1}}

\newcommand{\cA}{\mathcal{A}}

\newcommand{\cF}{\mathcal{F}}
\newcommand{\cG}{\mathcal{G}}

\newcommand{\cI}{\mathcal{I}}

\newcommand{\cN}{\mathcal{N}}
\newcommand{\cO}{\mathcal{O}}

\newcommand{\cR}{\mathcal{R}}

\newcommand{\cZ}{\mathcal{Z}}

\newcommand{\Nfour}{$\mathcal{N}=4$\xspace}

\newcommand{\NfSYMt}{\Nfour SYM theory\xspace}

\DeclareMathOperator{\T}{T}

\newlength{\eqoff}

\newcommand{\XYsize}{\scriptscriptstyle}

\newcommand{\Rem}{\cR^{(2)}}
\newcommand{\rem}{R^{(2)}}
\newcommand{\remi}{\rem_i}
\newcommand{\remif}[2]{(\remi)_{\XYsize #1}^{\XYsize #2}}

\newcommand{\interaction}{I}
\newcommand{\Interaction}{\cI}
\newcommand{\interactionr}{\underline{\interaction}}
\newcommand{\Interactionr}{\underline{\Interaction}}
\newcommand{\inttwo}[1][]{\interaction^{(2)}_{#1}}

\newcommand{\intone}[1][]{\interaction^{(1)}_{#1}}
\newcommand{\intoner}[1][]{\interactionr^{(1)}_{#1}}
\newcommand{\inttwoi}{\interaction^{(2)}_i}
\newcommand{\inttwoip}{\interaction^{(2)}_{i+1}}
\newcommand{\intoneone}{\interaction^{(1)}_1}
\newcommand{\intonei}{\interaction^{(1)}_i}
\newcommand{\intoneip}{\interaction^{(1)}_{i+1}}

\newcommand{\intoneir}{\interactionr^{(1)}_i}
\newcommand{\intoneipr}{\interactionr^{(1)}_{i+1}}
\newcommand{\inttwoif}[2]{(\inttwoi)_{\XYsize #1}^{\XYsize #2}}

\newcommand{\intoneonef}[2]{(\intoneone)_{\XYsize #1}^{\XYsize #2}}
\newcommand{\intoneif}[2]{(\intonei)_{\XYsize #1}^{\XYsize #2}}

\newcommand{\atreef}[2]{(A^{(0)})_{\XYsize #1}^{\XYsize #2}}
\newcommand{\Z}{\cZ}
\newcommand{\zone}{\cZ^{(1)}}
\newcommand{\zonei}{\zone_i}
\newcommand{\zoneip}{\zone_{i+1}}
\newcommand{\ztwo}{\cZ^{(2)}}
\newcommand{\ztwoi}{\ztwo_i}

\newcommand{\zoneif}[2]{(\zonei)_{\XYsize #1}^{\XYsize #2}}
\newcommand{\zoneipf}[2]{(\zoneip)_{\XYsize #1}^{\XYsize #2}}
\newcommand{\ztwoif}[2]{(\ztwoi)_{\XYsize #1}^{\XYsize #2}}

\newcommand{\symb}{\mathcal{S}}
\newcommand{\Dila}{\mathfrak{D}}
\newcommand{\dila}{\mathfrak{D}}
\newcommand{\dilatwoi}{\dila^{(2)}_i}
\newcommand{\dilatwo}[1][]{\dila^{(2)}_{#1}}
\newcommand{\dilatwoif}[2]{(\dilatwoi)_{\XYsize #1}^{\XYsize #2}}
\newcommand{\dilaone}[1][]{\dila^{(1)}_{#1}}
\newcommand{\dilaonei}{\dila^{(1)}_i}
\newcommand{\dilaoneif}[2]{(\dilaonei)_{\XYsize #1}^{\XYsize #2}}
\newcommand{\perm}{\mathbb{P}}

\newcommand{\peps}{\varepsilon}
\newcommand{\teps}{\epsilon}

\newcommand{\PSU}[1]{\operatorname{PSU}(#1)}
\renewcommand{\SU}[1]{\operatorname{SU}(#1)}
\renewcommand{\SO}[1]{\operatorname{SO}(#1)}

\newcommand{\sfrac}[2]{{\textstyle\frac{#1}{#2}}}
\numberwithin{equation}{section}

\usepackage{feynmp}
\makeatletter
\DeclareRobustCommand*{\bfseries}{%
  \not@math@alphabet\bfseries\mathbf
  \fontseries\bfdefault\selectfont
  \boldmath
}

\makeatother

\usepackage{lipsum}
\makeatletter
\if@titlepage
  \renewenvironment{abstract}{%
      \titlepage
      \null\vfil
      \@beginparpenalty\@lowpenalty
      \begin{center}%
        \bfseries \abstractname
        \@endparpenalty\@M
      \end{center}}%
     {\par\vfil\null\endtitlepage}
\else
  \renewenvironment{abstract}{%
      \if@twocolumn
        \section*{\abstractname}%
      \else
        \small
        \begin{center}%
          {\bfseries \abstractname\vspace{-.5em}\vspace{\z@}}%
        \end{center}%
        \quotation
      \fi}
      {\if@twocolumn\else\endquotation\fi}
\fi
\makeatother

\newcommand{\beq}{\begin{equation}}
\newcommand{\eeq}{\end{equation}}
\newcommand{\beqa}{\begin{eqnarray}}
\newcommand{\eeqa}{\end{eqnarray}}
\newcommand{\bea}{\begin{aligned}}
\newcommand{\eea}{\end{aligned}}

\title{On-Shell Methods for the Two-Loop Dilatation Operator and Finite Remainders}
\author{
Florian Loebbert, 
Dhritiman Nandan, 
Christoph Sieg,
Matthias Wilhelm, 
Gang Yang
}

\begin{document}
\begin{fmffile}{diagramsremainder}
\fmfcmd{%
thin := 1pt; 
thick := 2thin;
arrow_len := 3mm;
arrow_ang := 15;
curly_len := 3mm;
dash_len :=0.3; 
dot_len := 0.75mm; 
wiggly_len := 2mm; 
wiggly_slope := 60;
zigzag_len := 2mm;
zigzag_width := 2thick;
decor_size := 5mm;
dot_size := 2thick;
}
\fmfcmd{%
marksize=7mm;
def draw_cut(expr p,a) =
  begingroup
    save t,tip,dma,dmb; pair tip,dma,dmb;
    t=arctime a of p;
    tip =marksize*unitvector direction t of p;
    dma =marksize*unitvector direction t of p rotated -90;
    dmb =marksize*unitvector direction t of p rotated 90;
    linejoin:=beveled;
    drawoptions(dashed dashpattern(on 3bp off 3bp on 3bp));
    draw ((-.5dma.. -.5dmb) shifted point t of p);
    drawoptions();
  endgroup
enddef;
style_def phantom_cut expr p =
    save amid;
    amid=.5*arclength p;
    draw_cut(p, amid);
    draw p;
enddef;
}
\fmfcmd{%
smallmarksize=4mm;
def draw_smallcut(expr p,a) =
  begingroup
    save t,tip,dma,dmb; pair tip,dma,dmb;
    t=arctime a of p;
    tip =smallmarksize*unitvector direction t of p;
    dma =smallmarksize*unitvector direction t of p rotated -90;
    dmb =smallmarksize*unitvector direction t of p rotated 90;
    linejoin:=beveled;
    drawoptions(dashed dashpattern(on 2bp off 2bp on 2bp) withcolor black);
    draw ((-.5dma.. -.5dmb) shifted point t of p);
    drawoptions();
  endgroup
enddef;
style_def phantom_smallcut expr p =
    save amid;
    amid=.5*arclength p;
    draw_smallcut(p, amid);
    draw p;
enddef;
}
\fmfcmd{%
hugemarksize=11mm;
def draw_hugecut(expr p,a) =
  begingroup
    save t,tip,dma,dmb; pair tip,dma,dmb;
    t=arctime a of p;
    tip =hugemarksize*unitvector direction t of p;
    dma =hugemarksize*unitvector direction t of p rotated -90;
    dmb =hugemarksize*unitvector direction t of p rotated 90;
    linejoin:=beveled;
    drawoptions(dashed dashpattern(on 2bp off 2bp on 2bp) withcolor black);
    draw ((-.5dma.. -.5dmb) shifted point t of p);
    drawoptions();
  endgroup
enddef;
style_def phantom_hugecut expr p =
    save amid;
    amid=.5*arclength p;
    draw_hugecut(p, amid);
    draw p;
enddef;
}

\begingroup\parindent0pt
\begin{flushright}\footnotesize
\texttt{HU-MATH-2015-04}\\
\texttt{HU-EP-15/19} 
\end{flushright}
\vspace*{4em}
\centering
\begingroup\LARGE
\bf
On-Shell Methods for the Two-Loop Dilatation Operator and Finite Remainders
\par\endgroup
\vspace{3.5em}
\begingroup\normalsize\bf
Florian Loebbert$^{\text{a}}$, Dhritiman Nandan$^{\text{a,b}}$, Christoph Sieg$^{\text{a,b}}$, \\ Matthias Wilhelm$^{\text{a,b}}$, Gang Yang$^{\text{a}}$
\par\endgroup
\vspace{1em}
\begingroup\itshape
$^{\text{a}}$Institut für Physik\\
$^{\text{b}}$Institut für Mathematik\\
Humboldt-Universität zu Berlin\\
IRIS Gebäude, 
Zum Großen Windkanal 6,
12489 Berlin
\par\endgroup
\vspace{1em}
\begingroup\ttfamily
\{loebbert, dhritiman, csieg, mwilhelm, gang.yang\}@physik.hu-berlin.de \\
\par\endgroup
\vspace{3.5em}
\endgroup

\begin{abstract}

\noindent
We compute the two-loop minimal form factors of all operators in the $\SU{2}$ sector of planar \NfSYMt via on-shell unitarity methods.  From the UV divergence of this result, we obtain the two-loop dilatation operator in this sector.
Furthermore, we calculate the corresponding finite remainder functions. Since the operators break the supersymmetry, the remainder functions do not have the property of uniform transcendentality. However, the leading transcendentality part turns out to be universal and is identical to the corresponding BPS expression.
The remainder functions are shown to satisfy linear relations which can be explained by Ward identities of form factors following from R-symmetry.
\end{abstract}

\thispagestyle{empty}

\newpage
\hrule
\tableofcontents
\afterTocSpace
\hrule
\afterTocRuleSpace

\section{Introduction}
\label{sec: introduction}

In the last years, form factors in $\mathcal{N}=4$ super Yang--Mills (SYM) theory have received increasing attention, both at weak coupling \cite{vanNeerven:1985ja,Brandhuber:2010ad,Bork:2010wf,Brandhuber:2011tv,Bork:2011cj,Henn:2011by,Gehrmann:2011xn,Brandhuber:2012vm,Bork:2012tt,Engelund:2012re,Johansson:2012zv,Boels:2012ew,Penante:2014sza,Brandhuber:2014ica,Bork:2014eqa,Wilhelm:2014qua,Nandan:2014oga} and at strong coupling \cite{Alday:2007he,Maldacena:2010kp,Gao:2013dza}.
Containing both on-shell states and local composite operators, form factors provide a useful bridge between the purely on-shell amplitudes and the off-shell world of correlation functions.  
In particular, powerful computational methods developed in the context of scattering amplitudes can be applied to form factors and to other important physical quantities via form factors, such as the spectrum of anomalous dimensions of composite operators and their correlation functions. 
The form factor $\hat{\cF}_{\cO}$ is defined as the matrix element of a given local operator $\cO(x)$ between the vacuum $\ket{0}$ and an on-shell $n$-particle state $\bra{1,\dots,n}$, 
i.e.\
\begin{equation}\label{eq: form factor intro}
 \hat{\cF}_{\cO}(1,\dots,n;q)=\int \de^4 x \e^{-iqx}\bra{1,\dots,n}\cO(x)\ket{0}\eqndot
\end{equation}
A special class of form factors are the so-called minimal form factors, which contain as many external fields $n$ as there are fields in the operator, and which will be of particular interest for this paper.

Understanding the connection between form factors and the spectral problem of \NfSYMt  was recently pushed forward in  \cite{Wilhelm:2014qua,Nandan:2014oga}. 
In \cite{Wilhelm:2014qua}, form factors for generic operators were investigated. In particular, it was shown that the complete one-loop dilatation operator \cite{Beisert:2003jj} can be derived using one-loop minimal form factors, which explains the 
relation between the one-loop dilatation operator and the four-point scattering amplitude derived from symmetry in \cite{Zwiebel:2011bx}.%
\footnote{Moreover, in \cite{Zwiebel:2011bx} symmetry was used to show that all tree-level scattering amplitudes  
are related to certain contributions to the dilatation operator.
 The picture of  \cite{Zwiebel:2011bx} is equivalent to taking  cuts of form factors.
} 
 In \cite{Nandan:2014oga}, it was demonstrated that form factors can also be used to calculate anomalous dimensions at two-loop order by investigating the Konishi primary operator.
In these studies, on-shell amplitude techniques have played a major role, in particular the (generalised) unitarity method \cite{Bern:1994zx,Bern:1994cg, Britto:2004nc}. 
 In order to treat general operators, however, an extension of this method is required \cite{Nandan:2014oga}.  

Interesting on-shell approaches towards the computation of correlation functions and the dilatation operator were also applied in the following works: see \cite{Engelund:2012re,Brandhuber:2015boa} for the application of generalised unitarity, \cite{Engelund:2015cfa} for a spacetime version thereof, \cite{Koster:2014fva,Chicherin:2014uca}  for twistor techniques and \cite{Brandhuber:2014pta} for the application of MHV diagrams.

Computing form factors and correlators of non-protected local gauge-invariant operators 
requires renormalisation, which in general implies the mixing of these operators. 
This procedure singles out certain subsectors, which are closed under renormalisation and which transform under subalgebras of the full $\PSU{2,2|4}$ symmetry  \cite{Beisert:2003jj}. The simplest testing ground for studying the full renormalisation problem of $\mathcal{N}=4$ SYM theory is given by the so-called $\SU{2}$ sector. The operators in this sector are built out of two complex scalar fields $X$ and $Y$ transforming in the fundamental representation of $\SU{2}$, e.g.\
$X=\phi_{14}$ and
$Y=\phi_{24}$. In particular, the 
single-trace operators are of the form 
$\mathcal{O}_\text{bare}=\Tr(X^{k_1}Y^{k_2}X^{k_3}Y^{k_4}\cdots)$, where
$k_j\in\{0,1,2,\dots\}$. 
The renormalised operators of the interacting theory are
obtained from these bare operators via the 
mixing matrix $\Z$ as 
\begin{align}
\label{eq: z expansion}
\mathcal{O}_\text{ren}&=\mathcal{Z}\mathcal{O}_\text{bare}\eqncom 
&
\Z&=\idm+g^2 \zone +g^4 \ztwo +\cO(g^6)\eqndot
\end{align}
The study of this mixing problem has been of great importance for capturing the novel integrable structures appearing in planar $\mathcal{N}=4$ SYM theory at higher loop orders \cite{Beisert:2010jr}.
At one-loop order, the crucial observation introducing integrability to planar $\mathcal{N}=4$ SYM theory was that the anomalous dilatation operator defined as
\begin{equation}
\label{eq: dilatation operator definition}
\delta\mathfrak{D}=-\mu\frac{\de}{\de \mu}\log \cZ = 2\varepsilon g^2 \frac{\partial}{\partial g^2} \log \cZ=\sum_{\ell=1 }^\infty g^{2\ell}\Dila^{(\ell)}
\end{equation}
takes the form of the integrable Heisenberg spin-chain Hamiltonian  within the $\SU{2}$ sector \cite{Minahan:2002ve}.%
\footnote{In \cite{Minahan:2002ve}, the larger $\SO6$ sector was actually considered.}
The central role of the dilatation operator and its interpretation as an (asymptotic) spin-chain Hamiltonian was further emphasized in \cite{Beisert:2003tq}, where the two-loop  dilatation operator with $\SU{2}$ symmetry was computed from Feynman diagrams and its three-loop correction was derived under the assumption of integrability. A field-theoretic computation of the latter was later performed in \cite{Sieg:2010tz}.
Making use of integrability, a recursive construction for the asymptotic dilatation operator in the $\SU{2}$ sector is available by now, which allows to compute its operatorial form to high orders in the 't Hooft coupling constant \cite{Bargheer:2008jt,Bargheer:2009xy}.

In this paper, we continue the program of \cite{Wilhelm:2014qua,Nandan:2014oga} and study form factors and the dilatation operator at two-loop order in the full $\SU{2}$ sector.
We employ the unitarity method to obtain the complete two-loop form factors in this sector of planar $\mathcal{N}=4$ SYM theory. Interestingly, the form factor results satisfy linear relations. It turns out that they can be explained by Ward identities of form factors following from R-symmetry.

Form factors of non-protected operators contain both infrared (IR) divergences,  due to soft and collinear virtual momenta, and ultraviolet (UV) divergences.  
The information of the latter allows us to determine the renormalisation matrix $\cZ$, and therefore, the dilatation operator.%
\footnote{The anomalous dimensions can then be obtained as eigenvalues of the dilatation operator.} 
 In dimensional regularisation, where the four-dimensional theory is continued to $D=4-2\peps$ dimensions,%
\footnote{When continuing the spacetime dimension, also the fields have to be continued to $D=4-2\peps$. This leads to some important subtleties which have been analysed in detail in \cite{Nandan:2014oga}. These subtleties are, however, absent in the $\SU2$ sector.}
 all divergences are given by $1/\peps^k$ terms. In order to obtain the dilatation operator, we need to disentangle the IR and UV divergences, which is possible since the 
IR divergences have a well-understood universal structure \cite{Mueller:1979ih, Collins:1980ih, Sen:1981sd, Magnea:1990zb}.
Concretely, we will subtract the IR divergences via the BDS ansatz \cite{Anastasiou:2003kj,Bern:2005iz};\footnote{See also the previous studies of amplitudes in QCD \cite{Catani:1998bh, Sterman:2002qn}.} 
 a similar procedure has already been used in \cite{Wilhelm:2014qua,Nandan:2014oga}.

For amplitudes, it is well-known that the BDS ansatz does in general not give the full result but allows for a finite remainder function \cite{Alday:2007he}, which was first studied for the six-gluon case in \cite{Bartels:2008ce, Bern:2008ap,Drummond:2008aq}. For form factors of BPS operators, remainder functions have also been studied in \cite{Brandhuber:2012vm, Brandhuber:2014ica}. 
In particular, interesting properties associated to the so-called transcendentality were observed, such as the maximal transcendentality principle, which we will review below. 
In this paper, we will study the remainder functions of form factors of non-protected operators, where new features appear. 

Quantities in ${\cal N}=4$ SYM theory have shown interesting properties with respect to their transcendentality.  
Scattering amplitudes and form factors of BPS operators as well as their remainders have uniform transcendentality:%
\footnote{This is true at least in the cases of lower points or lower loops. There are known examples of amplitudes at sufficient high points in ${\cal N}=4$ SYM theory which are not given by transcendental functions but elliptic functions \cite{CaronHuot:2012ab,ArkaniHamed:2012nw,Nandan:2013ip}.} 
at $\ell$-loop order, they can be expressed as linear combinations of functions and numbers with transcendentality degree $2\ell$.
Furthermore, remarkable relations have been found between the results of ${\cal N}=4$ SYM theory and QCD. It was first argued in \cite{Kotikov:2001sc} that, for anomalous dimensions of twist-two operators, 
the ${\cal N}=4$ SYM theory result is given by  the leading transcendental part of the QCD result. This is usually referred to as \emph{the maximal transcendentality principle}; see also \cite{Kotikov:2004er, Kotikov:2006ts, Gehrmann:2011xn, Li:2014afw} for further discussions.  
While this heuristic relation was observed only for anomalous dimensions, 
in  \cite{Brandhuber:2012vm} it was found that the remainder function of certain BPS two-loop form factors matches exactly  the leading transcendental part of related two-loop Higgs-to-gluons amplitudes in QCD \cite{Gehrmann:2011aa}.  
This provides a first example where the maximal transcendentality principle is extended from pure numbers to functions which may have non-trivial kinematic dependence.\footnote{An interesting correspondence between the transcendental functions of ${\cal N}=4$ SYM theory and QCD was also found for  energy-energy correlations 
 \cite{Belitsky:2013ofa}.}

In this paper, we demonstrate that form factors of non-protected operators show new universality properties regarding their transcendentality. Since the considered operators break supersymmetry, the remainder functions are expected \emph{not} to have the property of \emph{uniform} transcendentality. 
However, we find that all contributions of maximal transcendentality are identical to the corresponding results of BPS form factors. This provides further evidence for the universality of the leading transcendental part, which furthermore has a non-trivial kinematic dependence. 

This paper is organised as follows. 
In section \ref{sec: form factors}, we present results for tree-level and one-loop form factors in the $\SU2$ sector. 
This also serves to introduce our conventions and notation.  
Moreover, we calculate the minimal two-loop form factors of such operators.
In section \ref{sec: remainder function and dilatation operator}, we extract the two-loop dilatation operator and two-loop remainder function from these results. 
Section \ref{sec: conclusion and outlook} contains our conclusions and outlook.
We provide simplified expressions for six-point amplitudes appearing in the unitarity calculation in appendix \ref{app: 6-point amplitudes}.

\section{Minimal form factors in the \texorpdfstring{$\SU{2}$}{SU(2)} sector}
\label{sec: form factors}

\subsection{Tree-level form factors}
\label{subsec: tree-level form factors}

In this subsection, we summarise some general facts about form factors and give explicit tree-level expressions that are required in the unitarity calculations of the subsequent subsections.

In analogy to amplitudes, we can strip off the gauge-group dependence of the form factors by introducing colour-ordered form factors $\cF_\cO$:
\begin{equation}\label{eq: def colour-ordered form factor}
 \begin{aligned}
 \hat{\cF}_{\cO}(1,\dots,n;q)&= \sum_{\sigma\in \SS_n/\ZZ_n} \Tr[\T^{a_{\sigma(1)}}\cdots\T^{a_{\sigma(n)}}] \cF_{\cO}(\sigma(1),\dots,\sigma(n);q) +\text{multi-trace terms} 
 \eqncom
 \end{aligned}
\end{equation}
where $\T^a$ with $a=1,\dots,N_c^2-1$ are the generators of the gauge group $\SU{N_c}$ and the sum is over all non-cyclic permutations. 
The multi-trace terms in \eqref{eq: def colour-ordered form factor} can start to appear at one-loop order but are suppressed in the planar limit, and will not be considered in this paper.

We describe the external on-shell states using Nair's $\cN=4$ on-shell superfield \cite{Nair:1988bq}:
\begin{equation}
\label{eq: N=4 superspace}
 \Phi(p,\eta)=g_+(p) +  \eta^A \, \bar\psi_A(p) + {\eta^A\eta^B \over 2!} \, \phi_{AB}(p) + { \teps_{ABCD} \eta^A\eta^B\eta^C \over 3!} \, \psi^D(p) + \eta^1\eta^2\eta^3\eta^4 \, g_-(p) \eqncom
\end{equation}
where $\eta_A$ are Graßmann variables that encode the flavour and helicity of the component
particles, and $A = 1, \dots, 4$ is the $\SU4$ R-symmetry index. 
In this formalism, we can combine form factors with different external fields into one super form factor. As we will see later, this also makes it easier to study the supersymmetry properties of the form factors.

In this paper, we focus on form factors in the $\SU{2}$ sector. The corresponding single-trace operators involve two complex scalar fields with a common $\SU4$ index, which are chosen explicitly as $X=\phi_{14}$ and $Y=\phi_{24}$. 
The tree-level minimal super form factor for the operator $\cO=\tr(XXYX\cdots)$ with $L=n$ fields, for instance, is simply given by
\begin{equation}
\label{eq:F-tree-su2}
  \cF^{(0)}_\cO(1,\dots,L;q)=\delta^4(q-\sum_{i=1}^L\lambda_i\lambdat_i)\left(\eta_1^1\eta_1^4\eta_2^1\eta_2^4\eta_3^2\eta_3^4\eta_4^1\eta_4^4\dots +\text{cyclic permutations}\right) \eqndot
\end{equation}
In general, the colour-ordered minimal tree-level super form factors of any operator can be obtained from the operator's oscillator representation by replacing the oscillators by spinor helicity variables and multiplying the result by the momentum-conserving delta function \cite{Wilhelm:2014qua}.

We also need the next-to-minimal tree-level form factors in the two-loop unitarity computation below, which contain one more external field than the minimal ones.
They may be computed easily by Feynman diagrams, or obtained from the BPS form factor component expressions, see e.g.\ \cite{Penante:2014sza}. 
For convenience, we provide some explicit rules that are useful in practice. There are four different cases that can occur.
In the first case, a $g_+$ can be inserted between two neighbouring positions $i$ and $i+1$. This leads to the following replacement in the colour-ordered minimal tree-level form factor:
\begin{equation}
 \cdots \eta_i^A\eta_i^B\eta_{i+1}^C\eta_{i+1}^D \cdots \longrightarrow \cdots \eta_i^A\eta_i^B \frac{\ab{i\,i\textrm{+}2}}{\ab{i\,i\textrm{+}1}\ab{i\textrm{+}1\,i\textrm{+}2}}  \eta_{i+2}^C\eta_{i+2}^D \cdots \eqndot
\end{equation}
In the second case, a $g_-$ can be inserted at the same position, leading to 
\begin{equation}
\cdots \eta_i^A\eta_i^B\eta_{i+1}^C\eta_{i+1}^D  \cdots \longrightarrow \cdots \eta_i^A\eta_i^B \frac{\sb{i\,i\textrm{+}2}}{\sb{i\,i\textrm{+}1}\sb{i\textrm{+}1\,i\textrm{+}2}} \eta_{i+1}^1\eta_{i+1}^2\eta_{i+1}^3\eta_{i+1}^4\eta_{i+2}^C\eta_{i+2}^D \cdots \eqndot
\end{equation}
In the third case, a $\phi_{CD}$ at position $i$ is split into two anti-fermions $\bar\psi_C$ and $\bar\psi_D$. This leads to 
\begin{equation}
\cdots \eta_{i-1}^A\eta_{i-1}^B \eta_i^C\eta_i^D  \eta_{i+1}^E\eta_{i+1}^F \cdots \longrightarrow \cdots \eta_{i-1}^A\eta_{i-1}^B \frac{1}{\ab{i\,i\textrm{+}1}} (\eta_i^C\eta_{i+1}^D - \eta_i^D\eta_{i+1}^C)   \eta_{i+2}^E\eta_{i+2}^F \cdots \eqndot
\end{equation}
In the fourth case, the $\phi_{CD}$ is split into two fermions $\psi^{C'}$ and $\psi^{D'}$ with $\teps_{CDC'D'}=1$, leading to 
\begin{equation}
\cdots \eta_{i-1}^A\eta_{i-1}^B \eta_i^C\eta_i^D  \eta_{i+1}^E\eta_{i+1}^F \cdots \longrightarrow \cdots \eta_{i-1}^A\eta_{i-1}^B \frac{-1}{\sb{i \, i\textrm{+}1}} (\bar\eta_{i,C'} \bar\eta_{i+1,D'} - \bar\eta_{i,D'} \bar\eta_{i+1,C'})   \eta_{i+2}^E\eta_{i+2}^F \cdots \eqncom
\end{equation}
where $\bar\eta_{i,A} = \frac{1}{3!}\teps_{ABCD}\eta_i^B\eta_i^C\eta_i^D$ and the minus sign is related to the order of the $\eta$'s.
The complete next-to-minimal form factor is obtained by summing over all four replacements and all insertion points.

\subsection{One-loop form factors}
\label{subsec: one-loop form factors}

In this subsection, we consider the one-loop minimal form factors in the $\SU{2}$ sector and show how to obtain the one-loop dilatation operator from them. 
This also allows us to introduce our notation and some important concepts that are required for the two-loop case. The results for the one-loop form factors, as well as the recipe to obtain the one-loop dilatation operator, were already given in \cite{Wilhelm:2014qua}. Here, a useful new formulation, given in \eqref{eq: one-loop int expansion}, is developed, which will be convenient to study the symmetry properties of form factors. 

Form factors in the loop expansion can be written in the following form:
\begin{equation}
\label{eq: loop correction}
 \cF_{\cO}=\big( 1+ g^2 \Interaction^{(1)}+g^4 \Interaction^{(2)}+ \dots\big) \cF_{\cO}^{(0)} \eqndot
\end{equation}
For operators that are eigenstates under renormalisation, such as BPS operators or the Konishi primary, $\Interaction^{(\ell)}$ is simply the ratio of the $\ell$-loop and tree-level form factor.
However, for form factors of operators that renormalise non-diagonally, this is no longer the case, because the loop corrections to vanishing tree-level form factors can be non-vanishing.
To overcome this problem, 
it is necessary to promote $\Interaction^{(\ell)}$ to an operator that acts on the tree-level form factor $\cF_{\cO}^{(0)}$ and creates a different tree-level form factor from it.

In the planar limit, connected $\ell$-loop interactions can maximally involve $\ell+1$ neighbouring fields in the colour-ordered form factor at a time. Hence, $\Interaction^{(\ell)}$ can be written as an interaction density that is summed over all insertion points. 
At one-loop order, the maximal interaction range is two, and we can write
\begin{equation}
\label{eq: one-loop interaction}
 \Interaction^{(1)}=\sum_{i=1}^L \interaction^{(1)}_{i\,i+1}.
\end{equation}
Here, $L$ denotes the length of the operator $\cO$, $\interaction^{(1)}_{i\,i+1}$ acts on the external fields $i$ and $i+1$ and cyclic identification $i+L\sim i$ is understood.
We depict $\interaction^{(1)}_{i\,i+1}$ as
\begin{equation}
 \interaction^{(1)}_{i\,i+1}=  
 \begin{aligned}
 \begin{tikzpicture}
  \drawvline{1}{1}
  \drawvline{2}{1}
  \drawtwoblob{1}{1}{$\intonei$}
 \end{tikzpicture}
 \end{aligned}
 \eqncom
\end{equation}
where we in general specify only the first field $i$ that is acted on 
when the range is explicitly specified by the number of occurring legs.
 
In the $\SU2$ sector, the following six range-two interactions are allowed by R-charge conservation:
$XX\rightarrow XX$, $XY\rightarrow XY$, $XY\rightarrow YX$, $YY\rightarrow YY$, $YX\rightarrow YX$ and $YX\rightarrow XY$. It is sufficient to consider the first three, as the last three can be obtained from them by replacing $X\leftrightarrow Y$, which is a symmetry of the theory.
We denote the contribution to a given combination of external fields $Z_A Z_B\rightarrow Z_C Z_D$ by $\intoneif{Z_AZ_B}{Z_CZ_D}$, where $Z_1=X$, $Z_2=Y$ and $A,B,C,D=1,2$.
In terms of these matrix elements, the operator 
$\interaction^{(1)}_{i\,i+1}$ 
is explicitly given by
\begin{equation}
\label{eq: one-loop int expansion}
 \interaction^{(1)}_{i\,i+1}=\sum_{A,B,C,D=1}^2 \intoneif{Z_AZ_B}{Z_CZ_D} \eta_{i}^C\frac{\partial}{\partial \eta_{i}^A} \eta_{i+1}^D\frac{\partial}{\partial \eta_{i+1}^B}\eqndot 
\end{equation}
\begin{figure}[t]
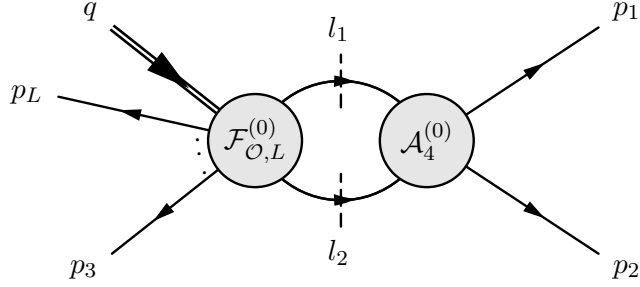

 \centering
$
\settoheight{\eqoff}{$\times$}%
\setlength{\eqoff}{0.5\eqoff}%
\addtolength{\eqoff}{-12.0\unitlength}%
\raisebox{\eqoff}{%
\fmfframe(2,2)(2,2){%
\begin{fmfchar*}(80,30)
\fmfleft{vp3,vp,vpL,vq}
\fmfright{vp2,vp1}
\fmf{dbl_plain_arrow,tension=1.2}{vq,v1}
\fmf{plain_arrow,tension=0}{v1,vpL}
\fmf{plain_arrow,tension=1.2}{v1,vp3}
\fmf{plain_arrow,left=0.7,label=$l_1\,,$,l.d=15}{v1,v2}
\fmf{plain_arrow,right=0.7,label=$l_2\,,$,l.d=15}{v1,v2}
\fmf{phantom_cut,left=0.7,tension=0}{v1,v2}
\fmf{phantom_cut,right=0.7,tension=0}{v1,v2}
\fmf{plain_arrow}{v2,vp1}
\fmf{plain_arrow}{v2,vp2}
\fmfv{decor.shape=circle,decor.filled=10,decor.size=35,label=$\cF_{\cO,,L}^{(0)}$,label.dist=0}{v1}
\fmfv{decor.shape=circle,decor.filled=10,decor.size=35,label=$\cA_{4}^{(0)}$,label.dist=0}{v2}
\fmffreeze
 \fmfcmd{pair vertq, vertpone, vertptwo, vertpthree, vertpL, vertone, verttwo; vertone = vloc(__v1); verttwo = vloc(__v2); vertq = vloc(__vq); vertpone = vloc(__vp1); vertptwo = vloc(__vp2); vertpthree = vloc(__vp3);vertpL = vloc(__vpL);}
 \fmfiv{label=$q$}{vertq}
 \fmfiv{label=$p_1$}{vertpone}
 \fmfiv{label=$p_2$}{vertptwo}
 \fmfiv{label=$p_3$}{vertpthree}
 \fmfiv{label=$p_L$}{vertpL}
 \fmfiv{label=$\cdot$,l.d=20,l.a=-150}{vertone}
 \fmfiv{label=$\cdot$,l.d=20,l.a=-165}{vertone}
 \fmfiv{label=$\cdot$,l.d=20,l.a=-180}{vertone}
\end{fmfchar*}%
}}%
$
\caption{The one-loop $(p_1+p_2)^2$ double cut.}
\label{fig: one-loop double cut}
\end{figure}

The matrix elements $\intoneif{Z_AZ_B}{Z_CZ_D}$ can be computed via unitarity. In the one-loop case, we only need to consider the double cut shown in figure \ref{fig: one-loop double cut}. Let us briefly consider the $\intoneonef{XY}{YX}$ case.
 The cut integrand is given by
\begin{equation}
\label{eq: double cut in example}
\int \de{\rm LIPS}(l_1, l_2) \de^4\eta_{l_1} \de^4\eta_{l_2} \cF^{(0)}_\cO(l_1^X, l_2^Y, p_3,\dots,p_L;q) \cA^{(0)}_4(-l_2, -l_1, p_1^Y, p_2^X) \eqncom
\end{equation}
where the tree-level form factor is given in \eqref{eq:F-tree-su2} and the four-point amplitude is given by the standard MHV expression. 
The labelling of the external legs with $X,Y$ in the tree-level amplitude and form factor means to take the corresponding $\eta$ components; for example, $\cA_4(-l_2, -l_1, p_1^Y, p_2^X)$ means to take the component of $\cA_4(-l_2, -l_1, p_1, p_2)$ containing the $(\eta_1^2 \eta_1^4)(\eta_2^1 \eta_2^4)$ factor.
Integrating out the $\eta_{l_i}$ variables, the cut integrand is given by\footnote{Note that
\begin{equation}
\cF^{(0)}_\cO(p_1^X,p_2^Y, p_3,\dots,p_L;q)\big|_{\eta_1^A=\eta_2^A=1}={\partial^2 \over \partial \eta_1^1 \partial \eta_1^4}{\partial^2 \over \partial \eta_2^2 \partial \eta_2^4}\cF^{(0)}_\cO(p_1,\dots,p_L;q)\Big|_{\eta_1^A=\eta_2^A=0}
\eqndot
\end{equation}}
\begin{equation}
\label{eq: simplified double cut in example}
(\eta_1^2 \eta_1^4)(\eta_2^1 \eta_2^4) \cF^{(0)}_\cO(p_1^X,p_2^Y, p_3,\dots,p_L;q)\big|_{\eta_1^A=\eta_2^A=1} \int \de{\rm LIPS}(l_1, l_2) \eqndot
\end{equation}
The variables $\eta_1$ and $\eta_2$ indicate that the result is not necessarily proportional to the tree-level form factor of the original operator but to the one of the operator in which the corresponding $X$ and $Y$ fields are permuted. 
The occurring phase space integral is simply the cut of a scalar bubble integral:
\begin{equation}
\FDinline[bubble,doublecut, cutlabels,twolabels,labelone=\scriptscriptstyle 1,labeltwo=\scriptscriptstyle 2] \eqndot
\end{equation}
At one-loop level, this cut is sufficient to determine the matrix element $\intoneonef{XY}{YX}$ as the bubble integral.
The other matrix elements can be obtained in a similar way.  
More details of such computations can be found e.g.\ in \cite{Wilhelm:2014qua,Nandan:2014oga}.

\begin{table}[t]
\begin{center}
\begin{tabular}{|l|c|c|c|}\hline
\qquad\qquad$\intoneif{}{}$&${}^{\XYsize XX}_{\XYsize XX}$&${}_{\XYsize XY}^{\XYsize XY}$&${}_{\XYsize XY}^{\XYsize YX}$
\\\hline
\FDinline[triangle,twolabels,labelone=\scriptscriptstyle i,labeltwo=\scriptscriptstyle i+1]
$s_{i\, i+1} $
&-1&-1&0
\\\hline
\FDinline[bubble,twolabels,labelone=\scriptscriptstyle i,labeltwo=\scriptscriptstyle i+1]
&0&-1&+1
\\\hline
\end{tabular}
\end{center}
\caption{Linear combinations of diagrams contributing to the minimal one-loop form factors in the $\SU{2}$ sector.}
\label{tab:1loop}
\end{table}

The one-loop results are summarised in table \ref{tab:1loop}. 
It is interesting to note that 
\begin{equation}
\label{eq: one-loop identity}
 \intoneif{XY}{XY}+\intoneif{XY}{YX}=\intoneif{XX}{XX}\eqndot
\end{equation}
This relation is a consequence of the $\SU2$ symmetry of the theory. Let us establish a formalism to deal with these symmetries in more detail since it demonstrates the general principle of how symmetries can be used to study form factors.

The $\PSU{2,2|4}$ symmetry of $\cN=4$ SYM theory leads to the following Ward identity of form factors:
\begin{equation}
\label{eq: Ward identity}
 \sum_{i=1}^n \mathfrak{J}_i^A \cF_\cO(1,\dots,n;q)= \cF_{\mathfrak{J}^A\cO}(1,\dots,n;q)\eqncom
\end{equation}
which holds for any generator $\mathfrak{J}_i^A$ of $\PSU{2,2|4}$; see e.g.\ \cite{Brandhuber:2011tv} for a detailed derivation. Let us consider explicitly the generators%
\footnote{In general, the generators of $\PSU{2,2|4}$ may obtain anomaly contributions, see e.g.\ \cite{Bargheer:2009qu}. These are, however, absent for $\SU2$.}
\begin{equation}
\mathfrak{J}_i^1=\eta_i^1\frac{\partial}{\partial\eta_i^2}+\eta_i^2\frac{\partial}{\partial\eta_i^1}\eqncom \quad
\mathfrak{J}_i^2=-i\eta_i^1\frac{\partial}{\partial\eta_i^2}+i\eta_i^2\frac{\partial}{\partial\eta_i^1}\eqncom \quad
\mathfrak{J}_i^3=\eta_i^1\frac{\partial}{\partial\eta_i^1}-\eta_i^2\frac{\partial}{\partial\eta_i^2}
\end{equation}
of $\SU2$.
Applying \eqref{eq: Ward identity} to \eqref{eq: loop correction} for the minimal tree-level and one-loop form factor, we find
\begin{equation}
\label{eq: symmetry commutator one-loop}
 [\mathfrak{J}^A,\Interaction^{(1)}]=0\eqncom
\end{equation}
where $\mathfrak{J}^A=\sum_{i=1}^L \mathfrak{J}_i^A$. Inserting \eqref{eq: one-loop int expansion} into \eqref{eq: symmetry commutator one-loop} yields \eqref{eq: one-loop identity} as well as similar identities.

The results of table~\ref{tab:1loop} contain the one-mass triangle and bubble integral, for which explicit
 expressions can be found e.g.\ in \cite{Smirnov:2004ym}.
The one-mass triangle integral is IR divergent and UV finite. The bubble integral, on the other hand, is IR finite but UV divergent.
Hence, the IR and UV divergences can be separated immediately.

The IR divergences of the above results match the universal form of one-loop IR divergences \cite{vanNeerven:1985ja}:
\begin{equation}
 \begin{aligned}
 \intone[i\,i+1] \Big|_{\text{IR}}&= -\frac{1}{\peps^2}(-s_{i\,i+1})^{-\peps} \idm_{i\,i+1} +\, \cO(\peps^0) \\
 &=\left[ -\frac{\gamma^{(1)}_{\text{cusp}}}{8\peps^2}-\frac{\cG_0^{(1)}}{4\peps}\right](-s_{i\,i+1})^{-\peps}\idm_{i\,i+1} +\, \cO(\peps^0)
  \eqncom
  \end{aligned}
\end{equation}
where $\gamma^{(1)}_{\text{cusp}}= 8$ is the one-loop cusp anomalous dimension and $\cG^{(1)}_0=0$ is the one-loop collinear anomalous dimension. We have also introduced the identity operator
\begin{equation}
\label{eq: def identity operator}
 \idm_{i\,i+1}= \sum_{A,B=1}^2\eta_i^A\frac{\partial}{\partial\eta_i^A}\eta_{i+1}^B\frac{\partial}{\partial\eta_{i+1}^B} \eqndot
\end{equation}

The UV divergences require the renormalisation of the operators.  The renormalised operators are defined in terms of the bare operators and the renormalisation constant $\cZ$ as shown in \eqref{eq: z expansion}. The renormalised form factor is nothing but the form factor of the renormalised operator.\footnote{This statement relies on the finiteness of $\mathcal{N}=4$ SYM theory and on a formulation in which also wave-function renormalisation is absent.} Since the form factor is linear in the operator, we can write in the case of the minimal form factor:
\begin{equation}
 \cF^{(0)}_{\cZ \cO}(1,\dots,L;q)= \cZ \cF^{(0)}_{\cO}(1,\dots,L;q)\eqncom
\end{equation}
where, on the right hand side, $\cZ$ acts as an \emph{operator} on the tree-level form factor, similar to $\Interaction^{(\ell)}$ discussed before, cf.\ \eqref{eq: loop correction}.

At one-loop level, $\cZ^{(1)}$ has to render the renormalised one-loop interaction
\begin{equation}
 \label{eq: renormalised one-loop interaction}
 \Interactionr^{(1)}=\Interaction^{(1)}+\cZ^{(1)}
\end{equation}
UV finite.
This means that $\zone_{i\,i+1}$ has to cancel the UV divergence of the bubble integrals occurring in $\interaction^{(1)}_{i\,i+1}$.
The UV divergence of the bubble integral is given by $\frac{1}{\peps}$.\footnote{We use a modified minimal subtraction scheme with effective planar coupling constant $g^2=\left(4\pi\e^{-\gamma_{\text{E}}}\right)^\varepsilon \frac{g_\YM^2 N_c}{(4\pi)^2}$.
}
Accordingly, using the results in table \ref{tab:1loop}, the one-loop renormalisation constant density is given by the matrix elements
\begin{equation}
\label{eq: one-loop renormalisation constant result}
 \zoneif{XX}{XX}=0\eqncom \qquad \zoneif{XY}{XY}=\frac{1}{\peps}\eqncom \qquad \zoneif{XY}{YX}=-\frac{1}{\peps}\eqndot
\end{equation}
It can be written in the compact operatorial form
\begin{equation}
 \mathcal{Z}_{i\,i+1}
 =
  \begin{aligned}
 \begin{tikzpicture}
  \drawvline{1}{1}
  \drawvline{2}{1}
  \drawtwoblob{1}{1}{$\zonei$}
 \end{tikzpicture}%
 \end{aligned} 
 =\frac{1}{\peps}(\idm-\PP)_{i\,i+1}
 \eqncom%
\end{equation}%
where $\idm$ is the identity operator \eqref{eq: def identity operator} and
\begin{equation}
\label{eq: def permutation operator}
 \PP_{i\,i+1}= \sum_{A,B=1}^2\eta_i^B\frac{\partial}{\partial\eta_i^A}\eta_{i+1}^A\frac{\partial}{\partial\eta_{i+1}^B}
\end{equation}
denotes the permutation operator.

In analogy to the renormalisation constant, we can also write the dilatation operator as an operator acting on the minimal tree-level form factor.
Applying \eqref{eq: dilatation operator definition} to \eqref{eq: one-loop renormalisation constant result}, we find the one-loop dilatation operator density
\begin{equation}
 \dilaoneif{XX}{XX}=0\eqncom \qquad \dilaoneif{XY}{XY}=2\eqncom \qquad \dilaoneif{XY}{YX}=-2\eqndot
\end{equation}
These expressions can be combined into the well-known form \cite{Minahan:2002ve}
\begin{equation}
 \dilaone[i\,i+1]=2(\idm-\PP)_{i\,i+1} \eqndot
\end{equation}
Let us now proceed to two-loop order.

\subsection{Two-loop form factors}
\label{subsec: two-loop form factors}

In the two-loop case, the range of connected interactions can be either two or three.  
Furthermore, two disconnected one-loop interactions can occur at two-loop level.
In total, we can introduce the two-loop operator $\Interaction^{(2)}$ similar to the one-loop case as
\begin{equation}
\label{eq: two-loop interaction}
 \Interaction^{(2)}=\sum_{i=1}^L \Big( \inttwo[i\,i+1\,i+2] + \inttwo[i\,i+1] + \frac{1}{2} \sum_{j=i+2}^{L+i-2} \intone[i\,i+1] \intone[j\,j+1] \Big) \eqncom
\end{equation}
where the last term accounts for the insertion of two one-loop interactions $\interaction^{(1)}_{i\,i+1}$ at non-overlapping positions.  The two-loop interactions $\interaction^{(2)}_{i\,i+1}$ and $\interaction^{(2)}_{i\,i+1\,i+2}$ are given by
\begin{equation}
\begin{aligned}
\label{eq: two-loop int expansion}
 \interaction^{(2)}_{i\,i+1} 
 & = \,\,\,\,\!\begin{aligned}
 \begin{tikzpicture}
  \drawvline{1}{1}
  \drawvline{2}{1}
  \drawtwoblob{1}{1}{$\inttwoi$}
 \end{tikzpicture}
 \end{aligned}\!\!\!\!\!\!\!\!\!\!\!\!
 && =\sum_{A,B,C,D=1}^2 \inttwoif{Z_AZ_B}{Z_CZ_D} \eta_{i}^C\frac{\partial}{\partial \eta_{i}^A} \eta_{i+1}^D\frac{\partial}{\partial \eta_{i+1}^B}\eqncom &&\\
  \interaction^{(2)}_{i\,i+1\,i+2} & =
 \, \begin{aligned}
 \begin{tikzpicture}
  \drawvline{1}{1}
  \drawvline{2}{1}
  \drawvline{3}{1}
  \drawthreeblob{1}{1}{$\inttwoi$}
 \end{tikzpicture}
 \end{aligned}\!\!\!\!
  && =\sum_{A,B,C,D,E,F=1}^2 \inttwoif{Z_AZ_BZ_C}{Z_DZ_EZ_F} \eta_{i}^D\frac{\partial}{\partial \eta_{i}^A} \eta_{i+1}^E\frac{\partial}{\partial \eta_{i+1}^B}  \eta_{i+2}^F\frac{\partial}{\partial \eta_{i+2}^C}\eqndot &&
  \end{aligned}
\end{equation}

For interaction range two, three distinct cases occur: $XX\rightarrow XX$, $XY\rightarrow XY$ and $XY\rightarrow YX$.
\begin{table}[tp]
\begin{center}
\begin{tabular}{|l|c|c|c|}\hline
\qquad\qquad$\inttwoif{}{}$&${}^{\XYsize XX}_{\XYsize XX}$&${}_{\XYsize XY}^{\XYsize XY}$&${}_{\XYsize XY}^{\XYsize YX}$
\\\hline
\FDinline[rainbow,twolabels,labelone=\scriptscriptstyle i,labeltwo=\scriptscriptstyle i+1]
$s_{i\,i+1}^2$
&+1&+1&0
\\
\FDinline[doubletriangletwo,threelabels,labelone=\scriptscriptstyle i,labeltwo=\scriptscriptstyle \phantom{i+1},labelthree=\scriptscriptstyle i+1]
$s_{i\,i+1}$
&+1&+1&0
\\\hline
\FDinline[rainbow,momentum,twolabels,labelone=\scriptscriptstyle i,labeltwo=\scriptscriptstyle i+1] 
$ s_{i\, i+1}s_{i\, l}$
&0&+1&-1
\\
\FDinline[fishtop,twolabels,labelone=\scriptscriptstyle i,labeltwo=\scriptscriptstyle i+1]
&0&+1&-1
\\
\FDinline[fishbottom,twolabels,labelone=\scriptscriptstyle i,labeltwo=\scriptscriptstyle i+1]
&0&+1&-1
\\\hline
\end{tabular}
\end{center}
\caption{Linear combinations of diagrams of range two contributing to the minimal two-loop form factors in the $\SU{2}$ sector. Terms between horizontal lines always occur in fixed combinations.}
\label{tab:2looprange2}
\end{table}
For interaction range three, six distinct cases occur: $XXX\rightarrow XXX$, $XXY\rightarrow XXY$, $XYX\rightarrow XYX$, $XXY\rightarrow XYX$, $XYX\rightarrow XXY$ and $XXY\rightarrow YXX$.
The remaining combinations can be obtained from these cases by exchanging $X\leftrightarrow Y$ and by using parity, i.e.\ reverting the order of the fields. 
\begin{table}[t]
\begin{center}
\begin{tabular}{|l|c|c|c|c|c|c|c|}\hline
  \qquad\qquad$\inttwoif{}{}$&${}_{\XYsize XXX}^{\XYsize XXX}$&${}_{\XYsize XXY}^{\XYsize XXY}$&${}_{\XYsize XYX}^{\XYsize XYX}$&${}_{\XYsize XXY}^{\XYsize XYX}$&${}_{\XYsize XYX}^{\XYsize XXY}$&${}_{\XYsize XXY}^{\XYsize YXX}$
\\\hline
\FDinline[trianglebox,momentum,threelabels,labelone=\scriptscriptstyle i,labeltwo=\scriptscriptstyle i+1,labelthree=\scriptscriptstyle i+2]
$s_{i\,l}s_{i+1\,i+2}$
&+1&+1&+1&0&0&0
\\
\FDinline[boxtriangle,momentum,threelabels,labelone=\scriptscriptstyle i,labeltwo=\scriptscriptstyle i+1,labelthree=\scriptscriptstyle i+2] 
$s_{i\,i+1}s_{i+2\,l}$
&+1&+1&+1&0&0&0
\\
\FDinline[doubletrianglethree,threelabels,labelone=\scriptscriptstyle i,labeltwo=\scriptscriptstyle i+1,labelthree=\scriptscriptstyle i+2]
$s_{i\,i+1\,i+2}$
&-1&-1&-1&0&0&0
\\\hline
\FDinline[boxbubble,threelabels,labelone=\scriptscriptstyle i,labeltwo=\scriptscriptstyle i+1,labelthree=\scriptscriptstyle i+2]
$s_{i\,i+1}$
&0&+1&+1&-1&-1&0
\\
\FDinline[triangleshiftedtriangle,momentum,threelabels,labelone=\scriptscriptstyle i,labeltwo=\scriptscriptstyle i+1,labelthree=\scriptscriptstyle i+2]
$s_{i\,l}$
&0&+1&+1&-1&-1&0
\\
\FDinline[trianglebubblethree,threelabels,labelone=\scriptscriptstyle i,labeltwo=\scriptscriptstyle i+1,labelthree=\scriptscriptstyle i+2]
&0&-1&-1&+1&+1&0
\\\hline
\FDinline[shiftedtrianglebubble,threelabels,labelone=\scriptscriptstyle i,labeltwo=\scriptscriptstyle i+1,labelthree=\scriptscriptstyle i+2]
&0&0&+1&-1&0&+1
\\\hline
\FDinline[bubblebox,threelabels,labelone=\scriptscriptstyle i,labeltwo=\scriptscriptstyle i+1,labelthree=\scriptscriptstyle i+2]
$s_{i+1\,i+2}$
&0&0&+1&0&0&0
\\
\FDinline[shiftedtriangletriangle,momentum,threelabels,labelone=\scriptscriptstyle i,labeltwo=\scriptscriptstyle i+1,labelthree=\scriptscriptstyle i+2]
$s_{i+2\,l}$
&0&0&+1&0&0&0
\\
\FDinline[bubbletrianglethree,threelabels,labelone=\scriptscriptstyle i,labeltwo=\scriptscriptstyle i+1,labelthree=\scriptscriptstyle i+2]
&0&0&-1&0&0&0
\\\hline
\FDinline[bubbleshiftedtriangle,threelabels,labelone=\scriptscriptstyle i,labeltwo=\scriptscriptstyle i+1,labelthree=\scriptscriptstyle i+2]
&0&0&+1&0&-1&0
\\\hline
\end{tabular}
\end{center}
\caption{Linear combinations of integrals of range three contributing to the minimal two-loop form factors in the $\SU{2}$ sector. The integrals are grouped such that those between two horizontal lines are always occurring in the same combination. The second and fourth group as well as the third and fifth group are related by parity.}
\label{tab:2looprange3}
\end{table}
We collect our results for the corresponding matrix elements 
 in \tabref{tab:2looprange2} and \tabref{tab:2looprange3}. 
The matrix elements $\inttwoif{XX}{XX}$ and $\inttwoif{XXX}{XXX}$ occur in the BPS case and were computed in \cite{Brandhuber:2014ica} using the unitarity method. The other matrix elements can be calculated in a similar but slightly more involved computation. 
Let us give a brief account of this computation.

\begin{figure}[t]
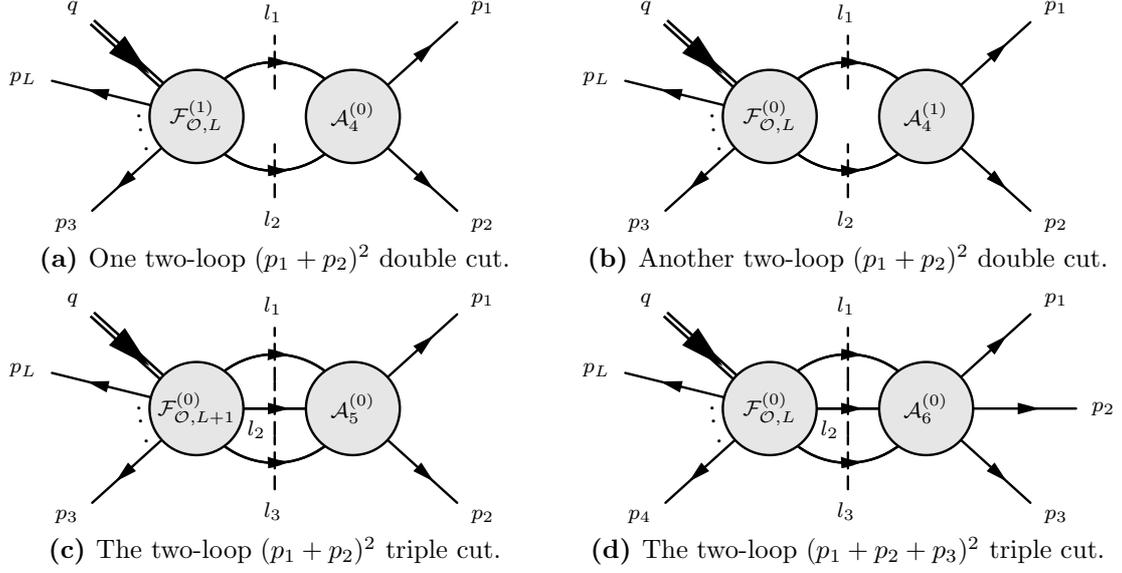
	
\centering
\begin{subfigure}[t]{0.49\textwidth}
 \centering
$
\settoheight{\eqoff}{$\times$}%
\setlength{\eqoff}{0.5\eqoff}%
\addtolength{\eqoff}{-12.0\unitlength}%
\raisebox{\eqoff}{%
\fmfframe(2,2)(2,2){%
\begin{fmfchar*}(60,25)
\fmfleft{vp3,vp,vpL,vq}
\fmfright{vp2,vp1}
\fmf{dbl_plain_arrow,tension=1.2}{vq,v1}
\fmf{plain_arrow,tension=0}{v1,vpL}
\fmf{plain_arrow,tension=1.2}{v1,vp3}
\fmf{plain_arrow,left=0.7,label=$\scriptstyle l_1\,,$,l.d=15,tension=0.8}{v1,v2}
\fmf{plain_arrow,right=0.7,label=$\scriptstyle l_2\,,$,l.d=15,tension=0.8}{v1,v2}
\fmf{phantom_cut,left=0.7,tension=0}{v1,v2}
\fmf{phantom_cut,right=0.7,tension=0}{v1,v2}
\fmf{plain_arrow,tension=1.2}{v2,vp1}
\fmf{plain_arrow,tension=1.2}{v2,vp2}
\fmfv{decor.shape=circle,decor.filled=10,decor.size=35,label=$\scriptstyle \cF_{\cO,,L}^{(1)}$,label.dist=0}{v1}
\fmfv{decor.shape=circle,decor.filled=10,decor.size=35,label=$\scriptstyle \cA_{4}^{(0)}$,label.dist=0}{v2}
\fmffreeze
 \fmfcmd{pair vertq, vertpone, vertptwo, vertpthree, vertpL, vertone, verttwo; vertone = vloc(__v1); verttwo = vloc(__v2); vertq = vloc(__vq); vertpone = vloc(__vp1); vertptwo = vloc(__vp2); vertpthree = vloc(__vp3);vertpL = vloc(__vpL);}
 \fmfiv{label=$\scriptstyle q$}{vertq}
 \fmfiv{label=$\scriptstyle p_1$}{vertpone}
 \fmfiv{label=$\scriptstyle p_2$}{vertptwo}
 \fmfiv{label=$\scriptstyle p_3$}{vertpthree}
 \fmfiv{label=$\scriptstyle p_L$}{vertpL}
 \fmfiv{label=$\cdot$,l.d=20,l.a=-150}{vertone}
 \fmfiv{label=$\cdot$,l.d=20,l.a=-165}{vertone}
 \fmfiv{label=$\cdot$,l.d=20,l.a=-180}{vertone}
\end{fmfchar*}%
}}%
$
\caption{One two-loop $(p_1+p_2)^2$ double cut.}
\label{fig: one two-loop double cut}
\end{subfigure}
\begin{subfigure}[t]{0.49\textwidth}
 \centering
$
\settoheight{\eqoff}{$\times$}%
\setlength{\eqoff}{0.5\eqoff}%
\addtolength{\eqoff}{-12.0\unitlength}%
\raisebox{\eqoff}{%
\fmfframe(2,2)(2,2){%
\begin{fmfchar*}(60,25)
\fmfleft{vp3,vp,vpL,vq}
\fmfright{vp2,vp1}
\fmf{dbl_plain_arrow,tension=1.2}{vq,v1}
\fmf{plain_arrow,tension=0}{v1,vpL}
\fmf{plain_arrow,tension=1.2}{v1,vp3}
\fmf{plain_arrow,left=0.7,label=$\scriptstyle l_1\,,$,l.d=15,tension=0.8}{v1,v2}
\fmf{plain_arrow,right=0.7,label=$\scriptstyle l_2\,,$,l.d=15,tension=0.8}{v1,v2}
\fmf{phantom_cut,left=0.7,tension=0}{v1,v2}
\fmf{phantom_cut,right=0.7,tension=0}{v1,v2}
\fmf{plain_arrow,tension=1.2}{v2,vp1}
\fmf{plain_arrow,tension=1.2}{v2,vp2}
\fmfv{decor.shape=circle,decor.filled=10,decor.size=35,label=$\scriptstyle \cF_{\cO,,L}^{(0)}$,label.dist=0}{v1}
\fmfv{decor.shape=circle,decor.filled=10,decor.size=35,label=$\scriptstyle \cA_{4}^{(1)}$,label.dist=0}{v2}
\fmffreeze
 \fmfcmd{pair vertq, vertpone, vertptwo, vertpthree, vertpL, vertone, verttwo; vertone = vloc(__v1); verttwo = vloc(__v2); vertq = vloc(__vq); vertpone = vloc(__vp1); vertptwo = vloc(__vp2); vertpthree = vloc(__vp3);vertpL = vloc(__vpL);}
 \fmfiv{label=$\scriptstyle q$}{vertq}
 \fmfiv{label=$\scriptstyle p_1$}{vertpone}
 \fmfiv{label=$\scriptstyle p_2$}{vertptwo}
 \fmfiv{label=$\scriptstyle p_3$}{vertpthree}
 \fmfiv{label=$\scriptstyle p_L$}{vertpL}
 \fmfiv{label=$\cdot$,l.d=20,l.a=-150}{vertone}
 \fmfiv{label=$\cdot$,l.d=20,l.a=-165}{vertone}
 \fmfiv{label=$\cdot$,l.d=20,l.a=-180}{vertone}
\end{fmfchar*}%
}}%
$
\caption{Another two-loop $(p_1+p_2)^2$ double cut.}
\label{fig: other two-loop double cut}
\end{subfigure}
\\[0.5\baselineskip]
\begin{subfigure}[t]{0.49\textwidth}
 \centering
$
\settoheight{\eqoff}{$\times$}%
\setlength{\eqoff}{0.5\eqoff}%
\addtolength{\eqoff}{-12.0\unitlength}%
\raisebox{\eqoff}{%
\fmfframe(2,2)(2,2){%
\begin{fmfchar*}(60,25)
\fmfleft{vp3,vp,vpL,vq}
\fmfright{vp2,vp1}
\fmf{dbl_plain_arrow,tension=1.2}{vq,v1}
\fmf{plain_arrow,tension=0}{v1,vpL}
\fmf{plain_arrow,tension=1.2}{v1,vp3}
\fmf{plain_arrow,left=0.7,label=$\scriptstyle l_1\,,$,l.d=15,tension=0.8}{v1,v2}
\fmf{plain_arrow,right=0.7,label=$\scriptstyle l_3\,,$,l.d=15,tension=0.8}{v1,v2}
\fmf{phantom_cut,left=0.7,tension=0}{v1,v2}
\fmf{phantom_cut,right=0.7,tension=0}{v1,v2}
\fmf{plain_arrow,tension=0,tag=1}{v1,v2}
\fmf{phantom_cut,tension=0}{v1,v2}
\fmf{plain_arrow,tension=1.2}{v2,vp1}
\fmf{plain_arrow,tension=1.2}{v2,vp2}
\fmfv{decor.shape=circle,decor.filled=10,decor.size=35,label=$\scriptstyle \cF_{\cO,,L+1}^{(0)}$,label.dist=0}{v1}
\fmfv{decor.shape=circle,decor.filled=10,decor.size=35,label=$\scriptstyle \cA_{5}^{(0)}$,label.dist=0}{v2}
\fmffreeze
 \fmfcmd{pair vertq, vertpone, vertptwo, vertpthree, vertpL, vertone, verttwo; vertone = vloc(__v1); verttwo = vloc(__v2); vertq = vloc(__vq); vertpone = vloc(__vp1); vertptwo = vloc(__vp2); vertpthree = vloc(__vp3);vertpL = vloc(__vpL);}
 \fmfiv{label=$\scriptstyle q$}{vertq}
 \fmfiv{label=$\scriptstyle p_1$}{vertpone}
 \fmfiv{label=$\scriptstyle p_2$}{vertptwo}
 \fmfiv{label=$\scriptstyle p_3$}{vertpthree}
 \fmfiv{label=$\scriptstyle p_L$}{vertpL}
 \fmfiv{label=$\cdot$,l.d=20,l.a=-150}{vertone}
 \fmfiv{label=$\cdot$,l.d=20,l.a=-165}{vertone}
 \fmfiv{label=$\cdot$,l.d=20,l.a=-180}{vertone}
\fmfipath{p[]}
\fmfiset{p1}{vpath1(__v1,__v2)}
\fmfiv{label=$\scriptstyle l_2\,,$,l.d=5,l.a=-115}{point length(p1)/2 of p1}
\end{fmfchar*}%
}}%
$
\caption{The two-loop $(p_1+p_2)^2$ triple cut.}
\label{fig: two-loop p1+p2 triple cut}
\end{subfigure}
\begin{subfigure}[t]{0.49\textwidth}
 \centering
$
\settoheight{\eqoff}{$\times$}%
\setlength{\eqoff}{0.5\eqoff}%
\addtolength{\eqoff}{-12.0\unitlength}%
\raisebox{\eqoff}{%
\fmfframe(2,2)(2,2){%
\begin{fmfchar*}(60,25)
\fmfleft{vp4,vp,vpL,vq}
\fmfright{vp3,vp2,vp1}
\fmf{dbl_plain_arrow,tension=1.2}{vq,v1}
\fmf{plain_arrow,tension=0}{v1,vpL}
\fmf{plain_arrow,tension=1.2}{v1,vp4}
\fmf{plain_arrow,left=0.7,label=$\scriptstyle l_1\,,$,l.d=15,tension=0.8}{v1,v2}
\fmf{plain_arrow,right=0.7,label=$\scriptstyle l_3\,,$,l.d=15,tension=0.8}{v1,v2}
\fmf{phantom_cut,left=0.7,tension=0}{v1,v2}
\fmf{phantom_cut,right=0.7,tension=0}{v1,v2}
\fmf{plain_arrow,tension=0,tag=1}{v1,v2}
\fmf{phantom_cut,tension=0}{v1,v2}
\fmf{plain_arrow,tension=1.2}{v2,vp1}
\fmf{plain_arrow,tension=0}{v2,vp2}
\fmf{plain_arrow,tension=1.2}{v2,vp3}
\fmfv{decor.shape=circle,decor.filled=10,decor.size=35,label=$\scriptstyle \cF_{\cO,,L}^{(0)}$,label.dist=0}{v1}
\fmfv{decor.shape=circle,decor.filled=10,decor.size=35,label=$\scriptstyle \cA_{6}^{(0)}$,label.dist=0}{v2}
\fmffreeze
 \fmfcmd{pair vertq, vertpone, vertptwo, vertpthree, vertpfour, vertpL, vertone, verttwo; vertone = vloc(__v1); verttwo = vloc(__v2); vertq = vloc(__vq); vertpone = vloc(__vp1); vertptwo = vloc(__vp2); vertpthree = vloc(__vp3); vertpfour = vloc(__vp4);vertpL = vloc(__vpL);}
 \fmfiv{label=$\scriptstyle q$}{vertq}
 \fmfiv{label=$\scriptstyle p_1$}{vertpone}
 \fmfiv{label=$\scriptstyle p_2$}{vertptwo}
 \fmfiv{label=$\scriptstyle p_3$}{vertpthree}
 \fmfiv{label=$\scriptstyle p_4$}{vertpfour}
 \fmfiv{label=$\scriptstyle p_L$}{vertpL}
 \fmfiv{label=$\cdot$,l.d=20,l.a=-150}{vertone}
 \fmfiv{label=$\cdot$,l.d=20,l.a=-165}{vertone}
 \fmfiv{label=$\cdot$,l.d=20,l.a=-180}{vertone}
\fmfipath{p[]}
\fmfiset{p1}{vpath1(__v1,__v2)}
\fmfiv{label=$\scriptstyle l_2\,,$,l.d=5,l.a=-115}{point length(p1)/2 of p1}
\end{fmfchar*}%
}}%
$
\caption{The two-loop $(p_1+p_2+p_3)^2$ triple cut.}
\label{fig: two-loop p1+p2+p3 triple cut}
\end{subfigure}
\caption{Unitary cuts of the minimal two-loop form factor.}
\label{fig: two-loop cuts}
\end{figure}

The cuts that have to be considered are depicted in figures \ref{fig: one two-loop double cut}, \ref{fig: other two-loop double cut}, \ref{fig: two-loop p1+p2 triple cut} and \ref{fig: two-loop p1+p2+p3 triple cut}.
 The tree-level next-to-minimal and one-loop minimal form factors, which occur as building blocks, are given in subsections \ref{subsec: tree-level form factors} and \ref{subsec: one-loop form factors}, respectively. 
The required tree-level and one-loop amplitudes are standard.
A particularly interesting cut is the triple cut shown in figure \ref{fig: two-loop p1+p2+p3 triple cut}, which involves the tree-level next-to-MHV six-point scalar amplitudes. As given explicitly in appendix \ref{app: 6-point amplitudes}, these scalar amplitudes take a simple form in terms of Mandelstam variables. 
Let us consider for example the $\inttwoif{XXY}{YXX}$ case. The cut integrand is given by the product of the minimal tree-level form factor and the six-point tree-level amplitude:
\begin{equation}
\label{eq:triple-cut-example}
\int \de{\rm LIPS}(l_1, l_2, l_3) \prod_{i=1}^3\de^4\eta_{l_i} \cF^{(0)}_\cO(l_1^X, l_2^X, l_3^Y, p_4,\dots,p_L;q) \cA_6(-l_3,-l_2, -l_1, p_1^Y, p_2^X, p_3^X) \eqndot
\end{equation}
After integrating out the $\eta_{l_i}$ variables, we obtain 
\begin{equation}
\label{eq: triple cut in example}
\begin{aligned}
\cF^{(0)}_\cO(p_1^X, p_2^X, p_3^Y,\dots,p_L;q)\big|_{\eta_1^A=\eta_2^A=\eta_3^A=1} (\eta_1^2\eta_1^4)(\eta_2^1\eta_2^4)(\eta_3^1\eta_3^4) \\
\qquad\qquad\times\int \de{\rm LIPS}(l_1, l_2, l_3) {1\over (-l_1+p_1+p_2)^2} \eqncom
\end{aligned}
\end{equation}
where we have used \eqref{eq: six- point amplitudes} for $\atreef{XXY}{YXX}$.\footnote{Note that there is a minus sign from the $\eta_{l_i}$ integration which cancels the sign in $\atreef{XXY}{YXX}$.} 
The phase space integral corresponds to the triple-cut loop integral
\begin{equation}
 \FDinline[shiftedtrianglebubble,alttriplecut,cutlabels,threelabels,labelone=\scriptscriptstyle 1,labeltwo=\scriptscriptstyle 2,labelthree=\scriptscriptstyle 3] \eqndot
\end{equation}
Similarly, 
each term in the other amplitudes in \eqref{eq: six- point amplitudes} is mapped to one graph in table \ref{tab:2looprange3} via the triple cut.

\def\middletension{0.8}

Looking at table \ref{tab:2looprange2} and table \ref{tab:2looprange3} and using the parity transformation, we observe some linear identities, e.g.\
\begin{equation}
 \inttwoif{XY}{XY}+\inttwoif{XY}{YX}=\inttwoif{XX}{XX}\eqncom
\end{equation}
and
\begin{equation}
\label{eq: I2 identities}
\begin{aligned}
 \inttwoif{XXY}{YXX}+\inttwoif{XXY}{XYX}+\inttwoif{XXY}{XXY}=\inttwoif{XXX}{XXX}\eqncom \\
 \inttwoif{XYX}{XYX}+\inttwoif{XYX}{YXX}+\inttwoif{XYX}{XXY}=\inttwoif{XXX}{XXX}\eqncom \\
 \inttwoif{XXY}{XYX}+\inttwoif{XXY}{YXX}=\inttwoif{XYX}{XXY}+\inttwoif{YXX}{XXY}\eqndot
\end{aligned}
\end{equation}
Similar to the one-loop case, all these identities are a consequence of $\SU2$ invariance and follow from the Ward identity \eqref{eq: Ward identity}, which at two-loop order yields
\begin{equation}
\label{eq: symmetry commutator two-loop}
 [\mathfrak{J}^A,\Interaction^{(2)}]=0\eqndot
\end{equation}

Given the full integrand of the two-loop form factor, we can perform a similar analysis as in the 
one-loop case.  
However, we will see that this requires a more involved subtraction of the IR divergences. 
This will be the topic of the next section.

\section{Two-loop dilatation operator and remainder function}
\label{sec: remainder function and dilatation operator}

In the one-loop case, the UV divergences stem from the bubble integrals alone. Therefore, the one-loop renormalisation constant can be read off directly from the coefficient of these integrals. This is no longer true for two-loop form factors, since the two-loop integrals in general contain a mixing of IR and UV divergences. However, IR divergences have a well-understood universal structure \cite{Mueller:1979ih, Collins:1980ih, Sen:1981sd, Magnea:1990zb}. This allows us to subtract the IR divergences systematically using the BDS ansatz \cite{Anastasiou:2003kj,Bern:2005iz}.

Similar to the one-loop case \eqref{eq: renormalised one-loop interaction}, the two-loop renormalised form factor 
is given by
\begin{equation}
 \label{eq: renormalised two-loop interaction}
 \Interactionr^{(2)}= \Interaction^{(2)}+\Interaction^{(1)}\cZ^{(1)}+\cZ^{(2)} \eqncom
\end{equation}
where
\begin{equation}
\label{eq: two-loop z}
 \ztwo=\sum_{i=1}^L \Big( \ztwo_{i\,i+1\,i+2} + \frac{1}{2} \sum_{j=i+2}^{L+i-2} \zone_{i\,i+1} \zone_{j\,j+1} \Big) \eqndot
\end{equation}
Applying the BDS ansatz \cite{Anastasiou:2003kj,Bern:2005iz} to the renormalised form factors, we obtain a finite two-loop remainder function: 
\begin{equation}
\label{eq: remainder}
 \Rem=\Interactionr^{(2)}(\peps) - \frac{1}{2}\left(\Interactionr^{(1)}(\peps)\right)^2 -  f^{(2)}(\peps)\Interactionr^{(1)}(2\peps)  
 + \cO (\peps ) \eqncom
 \end{equation}
where 
\begin{equation}
 f^{(2)} (\peps)= - 2 \zeta_2  - 2 \zeta_3 \peps - 2 \zeta_4 \peps^2 \eqndot 
\end{equation}

At two-loop order, connected interactions involve at most three fields of the composite operator, which have to be adjacent at the planar level. Hence, both the remainder function and the dilatation operator can be written in terms of densities that act only on triples of neighbouring sites at a time and are summed over all $L$ insertion points. 
For each triple of neighbouring points, we define the variables
\begin{equation}
\label{eq: uvw}
u_i = \frac{s_{i\,i+1}}{s_{i\,i+1 \, i+2}}\eqncom \quad
v_i = \frac{s_{i+1\,i+2}}{s_{i\,i+1 \, i+2}}\eqncom \quad
w_i = \frac{s_{i+2\,i}}{s_{i\,i+1 \, i+2}} \eqncom
\end{equation}
where 
\begin{equation}
s_{i\,i+1 \, i+2} = s_{i\,i+1}+s_{i+1\,i+2}+s_{i+2\,i} 
\end{equation}
and cyclic identification $i\sim i+L$ is understood. These variables satisfy $u_i+v_i+w_i = 1$.
The remainder $\Rem$ can be written in terms of its density as
\begin{equation}
\Rem 
= \sum_{i=1}^L \rem_{i\,i+1\,i+2} \eqndot
\end{equation}

An important subtlety arises due to the fact that the composite operators are not necessarily eigenstates under renormalisation. This requires a careful treatment of the product of one-loop form factors in  \eqref{eq: remainder}. As already mentioned, the renormalisation constant is a matrix (i.e.\ an operator) and so are the interactions. Hence, the one-loop product in \eqref{eq: remainder} should be understood as a product of operators. This can be explicitly depicted by the following equation in terms of graphs: 
\begin{equation}
\begin{aligned}
 \left(\Interactionr^{(1)}(\peps)\right)^2
 =
 \sum_{i=1}^L 
 \left(
 \sfrac{1}{2} 
 \begin{aligned}
 \begin{tikzpicture}
  \drawvline{1}{2}
  \drawvline{2}{2}
  \drawvline{3}{2}
  \drawtwoblob{1}{1}{$\intoneir$}
  \drawtwoblob{1}{2}{$\intoneir$}
 \end{tikzpicture}
 \end{aligned}
 +
 \begin{aligned}
 \begin{tikzpicture}
  \drawvline{1}{2}
  \drawvline{2}{2}
  \drawvline{3}{2}
  \drawtwoblob{1}{1}{$\intoneir$}
  \drawtwoblob{2}{2}{$\intoneipr$}
 \end{tikzpicture}
 \end{aligned}
 +
 \begin{aligned}
 \begin{tikzpicture}
  \drawvline{1}{2}
  \drawvline{2}{2}
  \drawvline{3}{2}
  \drawtwoblob{1}{2}{$\intoneir$}
  \drawtwoblob{2}{1}{$\intoneipr$}
 \end{tikzpicture}
 \end{aligned}
 + 
 \sfrac{1}{2} 
 \begin{aligned}
 \begin{tikzpicture}
  \drawvline{1}{2}
  \drawvline{2}{2}
  \drawvline{3}{2}
  \drawtwoblob{2}{1}{$\intoneipr$}
  \drawtwoblob{2}{2}{$\intoneipr$}
 \end{tikzpicture}
 \end{aligned}
 +  \sum_{j=i+2}^{L+i-2}
  \begin{aligned}
 \begin{tikzpicture}
  \drawvline{1}{1}
  \drawvline{2}{1}
  \drawtwoblob{1}{1}{$\intoneir$}
 \end{tikzpicture}
 \end{aligned}
  \begin{aligned}
 \begin{tikzpicture}
  \drawvline{1}{1}
  \drawvline{2}{1}
  \drawtwoblob{1}{1}{$\intoner[j]$}
 \end{tikzpicture}
 \end{aligned}
 \right)
 \eqndot
 \end{aligned}
\end{equation}
Note that the states corresponding to the internal lines are summed over as required for a product of operators.
The prefactors of $\frac{1}{2}$ stem from distributing products of densities with effective range two equally between the first two and the last two sites.

The remainder density, which itself is an operator, can be similarly expressed by the following graph equation:
\begin{equation}
\begin{aligned}
 &\rem_{i\,i+1\,i+2}=\\
&\qquad\quad\sfrac{1}{2} 
 \begin{aligned}
 \begin{tikzpicture}
  \drawvline{1}{1}
  \drawvline{2}{1}
  \drawvline{3}{1}
  \drawtwoblob{1}{1}{$\inttwoi$}
 \end{tikzpicture}
 \end{aligned}
 +
 \begin{aligned}
 \begin{tikzpicture}
  \drawvline{1}{1}
  \drawvline{2}{1}
  \drawvline{3}{1}
  \drawthreeblob{1}{1}{$\inttwoi$}
 \end{tikzpicture}
 \end{aligned}
 +
 \sfrac{1}{2} 
 \hspace{-1mm}
 \begin{aligned}
 \begin{tikzpicture}
  \drawvline{1}{1}
  \drawvline{2}{1}
  \drawvline{3}{1}
  \drawtwoblob{2}{1}{$\inttwoip$}
 \end{tikzpicture}
 \end{aligned}
 + 
 \begin{aligned}
 \begin{tikzpicture}
  \drawvline{1}{1}
  \drawvline{2}{1}
  \drawvline{3}{1}
  \drawthreeblob{1}{1}{$\ztwoi$}
 \end{tikzpicture}
 \end{aligned}
 +
 \sfrac{1}{2} 
  \hspace{-.5mm}
 \begin{aligned}
 \begin{tikzpicture}
  \drawvline{1}{2}
  \drawvline{2}{2}
  \drawvline{3}{2}
  \drawtwoblob{1}{1}{$\zonei$}
  \drawtwoblob{1}{2}{$\intonei$}
 \end{tikzpicture}
 \end{aligned}
 +
 \begin{aligned}
 \begin{tikzpicture}
  \drawvline{1}{2}
  \drawvline{2}{2}
  \drawvline{3}{2}
  \drawtwoblob{1}{1}{$\zonei$}
  \drawtwoblob{2}{2}{$\intoneip$}
 \end{tikzpicture}
 \end{aligned}
 +
 \begin{aligned}
 \begin{tikzpicture}
  \drawvline{1}{2}
  \drawvline{2}{2}
  \drawvline{3}{2}
  \drawtwoblob{1}{2}{$\intonei$}
  \drawtwoblob{2}{1}{$\zoneip$}
 \end{tikzpicture}
 \end{aligned}
 + 
 \sfrac{1}{2} 
  \hspace{-1.2mm}
 \begin{aligned}
 \begin{tikzpicture}
  \drawvline{1}{2}
  \drawvline{2}{2}
  \drawvline{3}{2}
  \drawtwoblob{2}{1}{$\zoneip$}
  \drawtwoblob{2}{2}{$\intoneip$}
 \end{tikzpicture}
 \end{aligned}\\
 &\qquad\quad-\sfrac{1}{2}
  \hspace{-1mm}
 \left(
  \sfrac{1}{2} 
   \hspace{-.5mm}
 \begin{aligned}
 \begin{tikzpicture}
  \drawvline{1}{2}
  \drawvline{2}{2}
  \drawvline{3}{2}
  \drawtwoblob{1}{1}{$\intoneir$}
  \drawtwoblob{1}{2}{$\intoneir$}
 \end{tikzpicture}
 \end{aligned}
 +
 \begin{aligned}
 \begin{tikzpicture}
  \drawvline{1}{2}
  \drawvline{2}{2}
  \drawvline{3}{2}
  \drawtwoblob{1}{1}{$\intoneir$}
  \drawtwoblob{2}{2}{$\intoneipr$}
 \end{tikzpicture}
 \end{aligned}
 +
 \begin{aligned}
 \begin{tikzpicture}
  \drawvline{1}{2}
  \drawvline{2}{2}
  \drawvline{3}{2}
  \drawtwoblob{1}{2}{$\intoneir$}
  \drawtwoblob{2}{1}{$\intoneipr$}
 \end{tikzpicture}
 \end{aligned}
 + 
 \sfrac{1}{2} 
  \hspace{-1mm}
 \begin{aligned}
 \begin{tikzpicture}
  \drawvline{1}{2}
  \drawvline{2}{2}
  \drawvline{3}{2}
  \drawtwoblob{2}{1}{$\intoneipr$}
  \drawtwoblob{2}{2}{$\intoneipr$}
 \end{tikzpicture}
 \end{aligned}
 \right)
 - f^{(2)}
 \left(
 \sfrac{1}{2} 
  \hspace{-.2mm}
 \begin{aligned}
 \begin{tikzpicture}
  \drawvline{1}{1}
  \drawvline{2}{1}
  \drawvline{3}{1}
  \drawtwoblob{1}{1}{$\intoneir$}
 \end{tikzpicture}
 \end{aligned}
 +
 \sfrac{1}{2} 
  \hspace{-1.5mm}
 \begin{aligned}
 \begin{tikzpicture}
  \drawvline{1}{1}
  \drawvline{2}{1}
  \drawvline{3}{1}
  \drawtwoblob{2}{1}{$\intoneipr$}
 \end{tikzpicture}
 \end{aligned}
 \right)_{\peps\rightarrow2\peps}
 \hspace{-2mm},
 \end{aligned}
 \label{eq: remainder density as graphs}
\end{equation}
where we have depicted the two-loop renormalisation constant density $\ztwo_{i\,i+1\,i+2}$ analogously to $\inttwo[i\,i+1\,i+2]$.
Requiring that this remainder density is finite allows us to fix the two-loop renormalisation constant density.

The integrals occurring in the two-loop result can be reduced to master integrals via IBP reduction, e.g.\ as implemented in the \texttt{Mathematica} package \texttt{LiteRed} \cite{Lee:2013mka}. The resulting master integrals can be found in \cite{Gehrmann:2000zt}.

\subsection{Renormalisation constant and dilatation operator}
\label{subsec: dilatation operator}

From the requirement that the two-loop renormalisation constant densities have to cancel all divergences in \eqref{eq: remainder density as graphs}, we find
\begin{equation}
\begin{aligned}
 \ztwoif{XXX}{XXX}&=0 \eqncom \qquad &&
 \ztwoif{XYX}{XYX}=+\frac{2}{ \peps ^2}-\frac{2}{ \peps } \eqncom \qquad &&
 \ztwoif{XXY}{XXY}=+\frac{1}{2 \peps ^2}-\frac{1}{2 \peps }\eqncom \\
 \ztwoif{XYX}{XXY}&=-\frac{1}{ \peps ^2}+\frac{1}{ \peps }  \eqncom\qquad &&
 \ztwoif{XXY}{XYX}=-\frac{1}{ \peps ^2}+\frac{1}{ \peps } \eqncom &&
 \ztwoif{XXY}{YXX}=+\frac{1}{2 \peps ^2}-\frac{1}{2 \peps } \eqndot 
 \end{aligned}
\end{equation}
Alternatively, this can be written in the operatorial form\footnote{Note that the coefficient of the simple pole coincides with the one of the double pole up to a sign.
This is a consequence of the fact that at two loops only one Feynman integral with overall UV divergence occurs in the $\SU2$ sector when using a manifestly IR finite formulation, and it does not hold for general operators; cf.\ \cite{Sieg:2010tz}. }
\begin{equation}
\label{eq: Z operatorial}
 \mathcal{Z}_{i\,i+1\,i+2}^{(2)}=
 \frac{1}{2}\left(\frac{1}{\peps ^2}-\frac{1}{\peps }\right)
 \big(\perm_{i\,i+1}\perm_{i+1\,i+2}+\perm_{i+1\,i+2}\perm_{i\,i+1}-3\perm_{i\,i+1}-3\perm_{i+1\,i+2}+4\big)\eqndot
\end{equation}

Using \eqref{eq: dilatation operator definition}, we have for the two-loop dilatation operator
\begin{equation}
\dilatwo = 4 \peps \Big(\mathcal{Z}^{(2)} - {1\over2} (\mathcal{Z}^{(1)} )^2 \Big) \eqncom
\end{equation}
where $(\mathcal{Z}^{(1)} )^2$ should be understood as a product of operators. For example,
 \begin{equation}
 \begin{aligned}
  \dilatwoif{XXY}{XYX} &=4\peps\left( \ztwoif{XXY}{XYX}  -\frac{1}{2} \zoneipf{XY}{YX}\zoneif{XX}{XX} - \frac{1}{2}\zoneif{XY}{XY}\zoneipf{XY}{YX} \right. 
  \\ 
   &\phantom{{}={}4\peps\left( \ztwoif{XXY}{XYX} \right. }  \left.   -\frac{1}{4} \zoneipf{XY}{YX}\zoneipf{XY}{XY} - \frac{1}{4}\zoneipf{YX}{YX}\zoneipf{XY}{YX} \right) = 4 \eqndot
  \end{aligned}
 \end{equation}
In total, we have 
\begin{equation}
\begin{aligned}
 \dilatwoif{XXX}{XXX}&=0 \eqncom \qquad &&
 \dilatwoif{XYX}{XYX}=-8 \eqncom \qquad &&
 \dilatwoif{XXY}{XXY}=-2 \eqncom  \\
 \dilatwoif{XYX}{XXY}&=4 \eqncom\qquad &&
 \dilatwoif{XXY}{XYX}=4 \eqncom \qquad &&
 \dilatwoif{XXY}{YXX}=-2 \eqncom
 \end{aligned}
\end{equation}
which agrees exactly with the known result \cite{Beisert:2003tq}
\begin{equation}
 \dilatwo[i\,i+1\,i+2]=-2\big(\perm_{i\,i+1}\perm_{i+1\,i+2}+\perm_{i+1\,i+2}\perm_{i\,i+1}-3\perm_{i\,i+1}-3\perm_{i+1\,i+2}+4\big)\eqndot
\end{equation}

\subsection{Finite remainders}

Next, we calculate the finite remainder densities.
The remainder densities fulfil analogous relations to \eqref{eq: I2 identities}:
\begin{equation}
\label{eq: remi identities}
\begin{aligned}
 \remif{XXY}{YXX}+\remif{XXY}{XYX}+\remif{XXY}{XXY}=\remif{XXX}{XXX}\eqncom \\
 \remif{XYX}{XYX}+\remif{XYX}{YXX}+\remif{XYX}{XXY}=\remif{XXX}{XXX}\eqncom \\
 \remif{XXY}{XYX}+\remif{XXY}{YXX}=\remif{XYX}{XXY}+\remif{YXX}{XXY}\eqndot
\end{aligned}
\end{equation}
These are equally a consequence of $\SU2$ symmetry and can be derived from 
\begin{equation}
\label{eq: symmetry commutator remainder}
 [\mathfrak{J}^A,\Rem]=0\eqncom
\end{equation}
which is a consequence of \eqref{eq: symmetry commutator one-loop} and \eqref{eq: symmetry commutator two-loop}.
Combining \eqref{eq: remi identities} with the symmetry under the exchange of $X\leftrightarrow Y$ and the reversion of the order of the fields, we can express all remainder densities in terms of $\remif{XXX}{XXX}$, $\remif{XXY}{XYX}$ and $\remif{XXY}{YXX}$.
Hence, it is enough to consider these three cases. 

The remainder density $\remif{XXX}{XXX}$ was already studied in \cite{Brandhuber:2014ica} and is of homogeneous transcendentality four:
\begin{equation}
 \remif{XXX}{XXX}=\remif{XXX}{XXX}\Big|_4 \eqncom
\end{equation}
which is given explicitly as 
 \begin{align}
 &\remif{XXX}{XXX}\Big|_4\nonumber\\\nonumber
 &=  -    \text{Li}_4(1-u_i)-\text{Li}_4(u_i)+\text{Li}_4\left(\frac{u_i - 1}{u_i}\right)
-  \log \left( \frac{1-u_i}{w_i }\right) 
\left[  \text{Li}_3\left(\frac{u_i - 1}{u_i}\right) - \text{Li}_3\left(1-u_i\right) \right] \\\nonumber
 &\phaneq- \log \left(u_i\right) \left[\text{Li}_3\left(\frac{v_i}{1-u_i}\right)+\text{Li}_3\left(-\frac{w_i}{v_i}\right) + \text{Li}_3\left(\frac{v_i-1}{v_i}\right)
 -\frac{1}{3}  \log ^3\left(v_i\right) -\frac{1}{3} \log ^3\left(1-u_i\right)  \right]
 \\\nonumber   
&\phaneq- \text{Li}_2\left(\frac{u_i-1}{u_i}\right) \text{Li}_2\left(\frac{v_i}{1-u_i}\right)+  \text{Li}_2\left(u_i\right) \left[
   \log \left(\frac{1-u_i}{ w_i }\right) \log \left(v_i \right) +\frac{1}{2} \log ^2\left( \frac{ 1-u_i }{ w_i }\right) \right] 
\\\nonumber
&\phaneq+  \frac{1}{24} \log ^4\left(u_i\right)
-\frac{1}{8} \log ^2\left(u_i\right) \log ^2\left(v_i\right)  - \frac{1}{2} \log ^2\left(1-u_i\right) \log \left(u_i\right) \log \left( \frac{ w_i }{ v_i}\right)\\\nonumber
&\phaneq- \frac{1}{2} \log \left(1-u_i\right) \log ^2\left(u_i\right) \log \left(v_i\right)
- \frac{1}{6} \log ^3\left(u_i\right) \log \left(w_i\right)  
  \\\nonumber
  &\phaneq- \zeta_2 \Big[ \log \left(u_i\right) \log \left(\frac{1-v_i}{ v_i} \right)
+ \frac{1}{2}\log ^2\left( \frac{ 1-u_i }{ w_i }\right) - \frac{1}{2}\log ^2\left(u_i\right)   \Big]
\\
&\phaneq + \zeta_3 \log (u_i) + \frac{\zeta_4}{ 2}  +G\left(\left\{1-u_i,1-u_i,1,0\right\},v_i\right) +  (u_i\, \leftrightarrow\, v_i)\eqndot
 \end{align}
Here, the Goncharov polylogarithm in the last line is the only piece that cannot be written in terms of classical polylogarithms.
This relatively compact expression was obtained using the symbol techniques  \cite{Goncharov09, Goncharov:2010jf}. The corresponding symbol is given by \cite{Brandhuber:2014ica}\footnote{The symbols can be conveniently calculated using the \texttt{Mathematica} code \cite{VerguCode}.}
 \begin{align} 
\symb\left( \remif{XXX}{XXX}\Big|_4 \right) &= 
- u_i\otimes (1-u_i)\otimes \left[ \frac{u_i-1}{u_i} \otimes \frac{v_i}{w_i}+ \frac{v_i}{w_i}\otimes \frac{w_i^2}{u_i v_i} \right]  - u_i\otimes u_i\otimes \frac{1-u_i}{v_i} \otimes \frac{w_i}{v_i}  \nonumber\\ 
&\phaneq
 - u_i\otimes v_i\otimes  \frac{v_i}{w_i} \otimes \frac{u_i}{w_i}
 - u_i\otimes v_i\otimes  \frac{u_i}{w_i} \otimes \frac{v_i}{w_i}
 + (u_i \leftrightarrow v_i ) \eqndot
\end{align}

The remainder density $\remif{XXY}{XYX}$ is of mixed transcendentality with degree ranging from three to zero.
Its contribution of degree three has the symbol 
\begin{align}
 \symb\left( \remif{XXY}{XYX}\Big|_3 \right)
&= v_i\otimes \frac{v_i}{1-v_i}\otimes \frac{u_i}{1-v_i} -v_i\otimes \frac{1-v_i}{u_i}\otimes \frac{v_i}{w_i}-u_i\otimes \frac{1-u_i}{v_i}\otimes \frac{v_i}{w_i}
\eqndot
\end{align}
The full transcendentality-three part can be given as
\begin{align}
\remif{XXY}{XYX}\Big|_3
&= \left[ \text{Li}_3\left(-\frac{u_i}{w_i}\right) - \log \left(u_i\right) \text{Li}_2\left(\frac{v_i}{1-u_i}\right)
+{1\over2} \log \left(1-u_i\right) \log \left(u_i\right) \log \left({w_i^2 \over 1-u_i} \right) \right.
\nonumber\\
&\phaneq \left. \phantom{\big[}- {1\over2} \text{Li}_3\left(-\frac{u_i v_i}{w_i}\right) - {1\over2}\log \left(u_i\right) \log \left(v_i\right) \log \left(w_i\right) - \frac{1}{12} \log ^3\left(w_i\right) + (u_i\, \leftrightarrow\, v_i) \right]
\nonumber\\
&\phaneq -\text{Li}_3\left(1-v_i\right)+\text{Li}_3\left(u_i\right)-\frac{1}{2} \log ^2\left(v_i\right) \log \left(\frac{1-v_i}{u_i}\right)
+\frac{1}{6} \pi ^2 \log \left(\frac{v_i}{w_i}\right) \nonumber\\
& \phaneq -\frac{1}{6} \pi ^2 \log \left(-s_{i\,i+1\,i+2}\right) 
 \eqndot
\end{align}
Together with the terms of lower transcendentality, we have
\begin{align}
 \remif{XXY}{XYX}&=\remif{XXY}{XYX}\Big|_3 +\text{Li}_2\left(1-u_i\right)+\text{Li}_2\left(1-v_i\right)\\
 &\phaneq+\log \left(u_i\right) \log \left(v_i\right)-\frac{1}{2} \log \left(-s_{i+1\,i+2}\right) \log \left(\frac{u_i}{v_i}\right)+2 \log \left(-s_{i\,i+1}\right) +\frac{\pi ^2}{3} -7
 \eqndot\nonumber
\end{align}
 
The final remainder density $\remif{XXY}{YXX}$ is of mixed transcendentality with degree ranging from two to zero. It reads
\begin{align}
 \remif{XXY}{YXX}
 &= \frac{1}{2} \log \left(-s_{i+1\,i+2}\right) \log \left(\frac{u_i}{v_i}\right)-\text{Li}_2\left(1-u_i\right)-\log \left(u_i\right) \log \left(v_i\right)+\frac{1}{2} \log ^2\left(v_i\right)\nonumber\\
 &\phaneq+\log \left(-s_{i+1\,i+2}\right)-2 \log \left(-s_{i\,i+1}\right)+\frac{7}{2} \eqndot
\end{align}
 
Let us emphasise that, if non-vanishing, the transcendentality-four contribution is the same for all remainder function densities.
Furthermore, there is only one transcendentality-three function and two functions of transcendentality smaller or equal to two that contribute to the results in the $\SU2$ sector. 
Notably, the highest degree of transcendentality $t=4-s$ is directly related to the shuffling number $s$ of the respective remainder density, i.e.\ to the number indicating by how many legs the field flavours are shuffled.
For instance, $\remif{XXY}{XYX}$ has shuffling number $s=1$ and maximal transcendentality degree $t=3$.

Interestingly, the rational pieces of the remainder function are connected to the dilatation operator as 
\begin{equation}
 \dilatwo[i\,i+1\,i+2]=-\frac{4}{7}\,\rem_{i\,i+1\,i+2}\Big|_0 \eqndot
\end{equation}

\section{Conclusion and outlook}
\label{sec: conclusion and outlook}

In this paper, we have calculated the two-loop minimal form factor for all operators in the $\SU{2}$ sector of planar $\mathcal{N}=4$ SYM theory via the on-shell method of unitarity. Moreover, we have extracted the corresponding two-loop remainder function and the two-loop dilatation operator from it. The results of this paper provide a solid stepping stone towards calculating the complete two-loop dilatation operator of $\mathcal{N}=4$ SYM theory. The employed method, however, is independent of the high symmetry of planar $\mathcal{N}=4$ SYM theory, in particular of its integrability, and it is thus also applicable to less symmetric theories.

The $\SU{2}$ sector is the simplest closed sector of the theory whose operators do not renormalise diagonally. 
It is hence well suited to study the occurrence of operator mixing and the dilatation operator.
Due to the on-shell nature of the external fields, the divergences of the form factors are a combination of UV and IR divergences.
We have disentangled the UV divergences from the IR divergences using the BDS ansatz and the universality of the latter.
Because of the operator mixing, we needed to promote the interactions to operators and the iterative structure of the BDS ansatz to an operatorial form as well. 
From the UV divergences, we have determined the renormalisation constants and the dilatation operator.

In contrast to the BPS case, the two-loop remainders of non-protected operators in the $\SU{2}$ sector do not exhibit maximal uniform transcendentality. However, their maximally transcendental part coincides with the remainder of the BPS vacuum computed in \cite{Brandhuber:2014ica}. 
These results and further evidence in other sectors lead us to conjecture that the two-loop remainder of every minimal form factor has the same degree-four part as the BPS one. 
Moreover, the two-loop remainder of the three-point form factor of every length-two operator should agree with the corresponding BPS remainder found in \cite{Brandhuber:2012vm}.
  It would be interesting to check our conjecture about this universality for a wider class of operators, or even to prove it.%
\footnote{One may expect that the maximally transcendental functions stem from the Laurent expansion of the functions with the highest-order poles in $\peps$, which originate from pure IR divergences. 
As the IR divergences are universal, the maximally transcendental part of the remainder should also be universal. 
However, one should be cautious about this argument: for example,  the leading transcendental functions of one-loop QCD amplitudes do not match with those of $\mathcal{N}=4$ SYM theory, although their IR divergences coincide \cite{Bern:1991aq}. 
An improved argument might proceed by showing the universality of the so-called leading singularities which are closely related to the maximally transcendental functions in the ${\de \log}$ form, as studied for amplitudes in $\mathcal{N}=4$ SYM theory e.g.\ in \cite{ArkaniHamed:2012nw}.}

Another observation is that the maximal transcendentality of the various remainders is related to their shuffling number, i.e.\ to the number indicating by how many places the field flavours are shuffled. It would be interesting to explore this pattern in larger sectors and at higher loops, where more complicated interactions contribute.

Soft or collinear limits of scattering amplitudes or form factors are typically given by lower-point amplitudes or form factors multiplied by a universal function. 
Thus, they provide important constraints and sometimes even allow to bootstrap the full function under consideration. 
As already observed in \cite{Brandhuber:2014ica} for the BPS case, soft and collinear limits of minimal form factors also do not vanish for the cases considered here. This may be surprising since a priori there is no physical interpretation for this limit because these form factors correspond to the \emph{minimal} physical configuration. Actually, similar questions also appear at the amplitude level. 
For example, taking certain soft or collinear limits of the six-scalar amplitudes given in appendix \ref{app: 6-point amplitudes} does not generate any physical amplitude, though the limit is non-zero. Via unitarity cuts, this consideration on the level of amplitudes affects the limits of minimal form factors.
It would be interesting to understand this point better, and to see if one can obtain the soft or collinear limit without computing the full quantity.

We have seen that the two-loop form factors, as well as the remainder functions and dilatation operator derived from them, obey Ward identities induced by the underlying R-symmetry. 
Going beyond the $\SU2$ sector, the realisation of similar Ward identities following from other symmetries should be more involved due to corrections to the generators which are absent for $\SU2$, cf.\ examples of such corrections in the case of spin chains \cite{Beisert:2003ys} or scattering amplitudes \cite{Bargheer:2009qu,Beisert:2010gn}. Capturing these extensions in the case of form factors is currently under investigation.

\section*{Acknowledgements}
It is a pleasure to thank Andreas Brandhuber, Burkhard Eden, Jan Fokken, Gregory Kor\-chemsky, Brenda Penante, Gabriele Travaglini and Christian Vergu for useful discussions. FL would like to thank Simon Caron-Huot for initial collaboration on a related project and for useful discussions.
We thank the Marie Curie network GATIS (gatis.desy.eu)
of the European Union's Seventh Framework Programme FP7/2007-2013/ under REA
Grant Agreement No 317089 for support.
MW dankt der Studienstiftung des deutschen Volkes f\"ur ein Pro\-mo\-tions\-f\"or\-der\-sti\-pen\-di\-um. 
DN and GY are supported by a DFG grant in the framework of the SFB 647 ``Raum-Zeit-Materie. Ana\-ly\-tische
und Geometrische Strukturen".

\appendix

\section{Six-point scalar amplitudes} 
\label{app: 6-point amplitudes}

In this appendix, we provide all six-point amplitudes that are required in the unitarity computation of the two-loop form factors in the $\SU2$ sector. We use the short notation
\begin{equation}
\atreef{Z_6 Z_5 Z_4}{Z_1 Z_2 Z_3} = A^{(0)}(1^{Z_1}, 2^{Z_2}, 3^{Z_3}, 4^{\bar Z_4}, 5^{\bar Z_5}, 6^{\bar Z_6})\big|_{\eta_i^A=1} \eqncom
\end{equation}
where, on the right-hand side, all momenta are taken to be outgoing and we have set all $\eta_i$ variables to $1$.

The required amplitudes can be explicitly given in terms of Mandelstam variables as
\begin{equation}
\label{eq: six- point amplitudes}
\begin{aligned}
\atreef{YXX}{XXY} = & - {1\over s_{234}} \eqncom \\
\atreef{XXY}{YXX} = & - {1\over s_{345}} \eqncom \\
\atreef{XXX}{XXX} = & {s_{23} s_{56} \over s_{16} s_{34} s_{234}} + {s_{12} s_{45} \over s_{16} s_{34} s_{345}} - {s_{123} \over s_{16} s_{34}} \eqncom \\
\atreef{XXY}{XYX} = & {s_{12} \over s_{16} s_{345}} + {s_{56} \over s_{16} s_{234}} - {1\over s_{16}} + {1\over s_{345}} \eqncom \\
\atreef{YXX}{XYX} = & {s_{23} \over s_{34} s_{234}} + {s_{45} \over s_{34} s_{345}} - {1\over s_{34}} + {1\over s_{234}} \eqncom \\
\atreef{XYX}{XXY} = & {s_{12} \over s_{16} s_{345}} + {s_{56} \over s_{16} s_{234}} - {1\over s_{16}} + {1\over s_{234}} \eqncom \\
\atreef{XYX}{YXX} = & {s_{23} \over s_{34} s_{234}} + {s_{45} \over s_{34} s_{345}} - {1\over s_{34}} + {1\over s_{345}} \eqncom \\
\atreef{XXY}{XXY} = & - {s_{23} s_{56} \over s_{16} s_{34} s_{234}} - {s_{12} s_{45} \over s_{16} s_{34} s_{345}} +  {s_{123} \over s_{16} s_{34}} - {s_{12} \over s_{16} s_{345}} - {s_{56} \over s_{16} s_{234}} + {1\over s_{16}} \eqncom \\
\atreef{YXX}{YXX} = & - {s_{23} s_{56} \over s_{16} s_{34} s_{234}} - {s_{12} s_{45} \over s_{16} s_{34} s_{345}} +  {s_{123} \over s_{16} s_{34}} - {s_{23} \over s_{34} s_{234}} - {s_{45} \over s_{34} s_{345}} + {1\over s_{34}} \eqncom \\
\atreef{XYX}{XYX} = & - {s_{23} s_{56} \over s_{16} s_{34} s_{234}} - {s_{12} s_{45} \over s_{16} s_{34} s_{345}} +  {s_{123} \over s_{16} s_{34}} \\ 
& - {s_{12} \over s_{16} s_{345}} - {s_{56} \over s_{16} s_{234}} + {1\over s_{16}} - {1\over s_{345}} - {s_{23} \over s_{34} s_{234}} - {s_{45} \over s_{34} s_{345}} + {1\over s_{34}} - {1\over s_{234} } \eqncom 
\end{aligned}
\end{equation}
where all poles are physical.\footnote{The BCFW recursion relation or MHV rule methods directly give results in a much more complicated form, usually involving spurious poles. We have checked that these different methods give results equivalent to \eqref{eq: six- point amplitudes}.} Through the triple cut shown in figure \ref{fig: two-loop p1+p2+p3 triple cut}, each term in \eqref{eq: six- point amplitudes} is exactly mapped to one graph in table \ref{tab:2looprange3}.

It is easy to find various relations among these amplitudes, such as
\begin{equation}
\begin{aligned}
& \atreef{XXY}{YXX} + \atreef{XXY}{XYX} +  \atreef{XXY}{XXY}  = - \atreef{XXX}{XXX} \eqncom \\
& \atreef{XYX}{XYX} + \atreef{XYX}{YXX} + \atreef{XYX}{XXY}  = - \atreef{XXX}{XXX} \eqncom \\
& \atreef{XXY}{XYX} + \atreef{XXY}{YXX} = \atreef{XYX}{XXY} + \atreef{YXX}{XXY}  \eqndot 
\end{aligned}
\end{equation}
These are the counterparts of \eqref{eq: I2 identities} and nothing but supersymmetric Ward identities (SWI) for amplitudes \cite{Grisaru:1976vm}.%
\footnote{The minus sign in front of $\atreef{XXX}{XXX}$ is due to the convention for the fermionic $\eta$ variable in amplitudes.}

\bibliographystyle{utcaps}
\bibliography{ThesisINSPIRE}

\providecommand{\href}[2]{#2}\begingroup\raggedright\begin{thebibliography}{10}

\bibitem{vanNeerven:1985ja}
W.~van Neerven, ``{Infrared Behavior of On-shell Form-factors in a $\cN=4$
  Supersymmetric {Yang-Mills} Field Theory},''
\href{http://dx.doi.org/10.1007/BF01571808}{{\em Z.Phys.} {\bfseries C30}
  (1986) 595}.

\bibitem{Brandhuber:2010ad}
A.~Brandhuber, B.~Spence, G.~Travaglini, and G.~Yang, ``{Form Factors in
  $\cN=4$ Super Yang-Mills and Periodic Wilson Loops},''
  \href{http://dx.doi.org/10.1007/JHEP01(2011)134}{{\em JHEP} {\bfseries 1101}
  (2011) 134},
\href{http://arxiv.org/abs/1011.1899}{{\ttfamily arXiv:1011.1899 [hep-th]}}.

\bibitem{Bork:2010wf}
L.~Bork, D.~Kazakov, and G.~Vartanov, ``{On form factors in $\cN=4$ SYM},''
  \href{http://dx.doi.org/10.1007/JHEP02(2011)063}{{\em JHEP} {\bfseries 1102}
  (2011) 063},
\href{http://arxiv.org/abs/1011.2440}{{\ttfamily arXiv:1011.2440 [hep-th]}}.

\bibitem{Brandhuber:2011tv}
A.~Brandhuber, O.~Gurdogan, R.~Mooney, G.~Travaglini, and G.~Yang, ``{Harmony
  of Super Form Factors},''
  \href{http://dx.doi.org/10.1007/JHEP10(2011)046}{{\em JHEP} {\bfseries 1110}
  (2011) 046},
\href{http://arxiv.org/abs/1107.5067}{{\ttfamily arXiv:1107.5067 [hep-th]}}.

\bibitem{Bork:2011cj}
L.~Bork, D.~Kazakov, and G.~Vartanov, ``{On MHV Form Factors in Superspace for
  $\mathcal{N}=4$ SYM Theory},''
  \href{http://dx.doi.org/10.1007/JHEP10(2011)133}{{\em JHEP} {\bfseries 1110}
  (2011) 133},
\href{http://arxiv.org/abs/1107.5551}{{\ttfamily arXiv:1107.5551 [hep-th]}}.

\bibitem{Henn:2011by}
J.~M. Henn, S.~Moch, and S.~G. Naculich, ``{Form factors and scattering
  amplitudes in $\cN=4$ SYM in dimensional and massive regularizations},''
  \href{http://dx.doi.org/10.1007/JHEP12(2011)024}{{\em JHEP} {\bfseries 1112}
  (2011) 024},
\href{http://arxiv.org/abs/1109.5057}{{\ttfamily arXiv:1109.5057 [hep-th]}}.

\bibitem{Gehrmann:2011xn}
T.~Gehrmann, J.~M. Henn, and T.~Huber, ``{The three-loop form factor in $\cN=4$
  super Yang-Mills},'' \href{http://dx.doi.org/10.1007/JHEP03(2012)101}{{\em
  JHEP} {\bfseries 1203} (2012) 101},
\href{http://arxiv.org/abs/1112.4524}{{\ttfamily arXiv:1112.4524 [hep-th]}}.

\bibitem{Brandhuber:2012vm}
A.~Brandhuber, G.~Travaglini, and G.~Yang, ``{Analytic two-loop form factors in
  $\cN=4$ SYM},'' \href{http://dx.doi.org/10.1007/JHEP05(2012)082}{{\em JHEP}
  {\bfseries 1205} (2012) 082},
\href{http://arxiv.org/abs/1201.4170}{{\ttfamily arXiv:1201.4170 [hep-th]}}.

\bibitem{Bork:2012tt}
L.~Bork, ``{On NMHV form factors in $\cN=4$ SYM theory from generalized
  unitarity},'' \href{http://dx.doi.org/10.1007/JHEP01(2013)049}{{\em JHEP}
  {\bfseries 1301} (2013) 049},
\href{http://arxiv.org/abs/1203.2596}{{\ttfamily arXiv:1203.2596 [hep-th]}}.

\bibitem{Engelund:2012re}
O.~T. Engelund and R.~Roiban, ``{Correlation functions of local composite
  operators from generalized unitarity},''
  \href{http://dx.doi.org/10.1007/JHEP03(2013)172}{{\em JHEP} {\bfseries 1303}
  (2013) 172},
\href{http://arxiv.org/abs/1209.0227}{{\ttfamily arXiv:1209.0227 [hep-th]}}.

\bibitem{Johansson:2012zv}
H.~Johansson, D.~A. Kosower, and K.~J. Larsen, ``{Two-Loop Maximal Unitarity
  with External Masses},''
  \href{http://dx.doi.org/10.1103/PhysRevD.87.025030}{{\em Phys.Rev.}
  {\bfseries D87} (2013) 025030},
\href{http://arxiv.org/abs/1208.1754}{{\ttfamily arXiv:1208.1754 [hep-th]}}.

\bibitem{Boels:2012ew}
R.~H. Boels, B.~A. Kniehl, O.~V. Tarasov, and G.~Yang, ``{Color-kinematic
  Duality for Form Factors},''
  \href{http://dx.doi.org/10.1007/JHEP02(2013)063}{{\em JHEP} {\bfseries 1302}
  (2013) 063},
\href{http://arxiv.org/abs/1211.7028}{{\ttfamily arXiv:1211.7028 [hep-th]}}.

\bibitem{Penante:2014sza}
B.~Penante, B.~Spence, G.~Travaglini, and C.~Wen, ``{On super form factors of
  half-BPS operators in $\mathcal{N}=4$ super Yang-Mills},''
  \href{http://dx.doi.org/10.1007/JHEP04(2014)083}{{\em JHEP} {\bfseries 1404}
  (2014) 083},
\href{http://arxiv.org/abs/1402.1300}{{\ttfamily arXiv:1402.1300 [hep-th]}}.

\bibitem{Brandhuber:2014ica}
A.~Brandhuber, B.~Penante, G.~Travaglini, and C.~Wen, ``{The last of the simple
  remainders},'' \href{http://dx.doi.org/10.1007/JHEP08(2014)100}{{\em JHEP}
  {\bfseries 1408} (2014) 100},
\href{http://arxiv.org/abs/1406.1443}{{\ttfamily arXiv:1406.1443 [hep-th]}}.

\bibitem{Bork:2014eqa}
L.~Bork, ``{On form factors in $ \mathcal{N}=4 $ SYM theory and polytopes},''
  \href{http://dx.doi.org/10.1007/JHEP12(2014)111}{{\em JHEP} {\bfseries 1412}
  (2014) 111},
\href{http://arxiv.org/abs/1407.5568}{{\ttfamily arXiv:1407.5568 [hep-th]}}.

\bibitem{Wilhelm:2014qua}
M.~Wilhelm, ``{Amplitudes, Form Factors and the Dilatation Operator in
  $\mathcal{N}=4$ SYM Theory},''
  \href{http://dx.doi.org/10.1007/JHEP02(2015)149}{{\em JHEP} {\bfseries 1502}
  (2015) 149},
\href{http://arxiv.org/abs/1410.6309}{{\ttfamily arXiv:1410.6309 [hep-th]}}.

\bibitem{Nandan:2014oga}
D.~Nandan, C.~Sieg, M.~Wilhelm, and G.~Yang, ``{Cutting through form factors
  and cross sections of non-protected operators in $\mathcal{N}=4 $ SYM},''
  \href{http://dx.doi.org/10.1007/JHEP06(2015)156}{{\em JHEP} {\bfseries 06}
  (2015) 156},
\href{http://arxiv.org/abs/1410.8485}{{\ttfamily arXiv:1410.8485 [hep-th]}}.

\bibitem{Alday:2007he}
L.~F. Alday and J.~Maldacena, ``{Comments on gluon scattering amplitudes via
  AdS/CFT},'' \href{http://dx.doi.org/10.1088/1126-6708/2007/11/068}{{\em JHEP}
  {\bfseries 0711} (2007) 068},
\href{http://arxiv.org/abs/0710.1060}{{\ttfamily arXiv:0710.1060 [hep-th]}}.

\bibitem{Maldacena:2010kp}
J.~Maldacena and A.~Zhiboedov, ``{Form factors at strong coupling via a
  Y-system},'' \href{http://dx.doi.org/10.1007/JHEP11(2010)104}{{\em JHEP}
  {\bfseries 1011} (2010) 104},
\href{http://arxiv.org/abs/1009.1139}{{\ttfamily arXiv:1009.1139 [hep-th]}}.

\bibitem{Gao:2013dza}
Z.~Gao and G.~Yang, ``{Y-system for form factors at strong coupling in $AdS_5$
  and with multi-operator insertions in $AdS_3$},''
  \href{http://dx.doi.org/10.1007/JHEP06(2013)105}{{\em JHEP} {\bfseries 1306}
  (2013) 105},
\href{http://arxiv.org/abs/1303.2668}{{\ttfamily arXiv:1303.2668 [hep-th]}}.

\bibitem{Beisert:2003jj}
N.~Beisert, ``{The complete one-loop dilatation operator of $\cN=4$
  superYang-Mills theory},''
  \href{http://dx.doi.org/10.1016/j.nuclphysb.2003.10.019}{{\em Nucl.Phys.}
  {\bfseries B676} (2004) 3--42},
\href{http://arxiv.org/abs/hep-th/0307015}{{\ttfamily arXiv:hep-th/0307015
  [hep-th]}}.

\bibitem{Zwiebel:2011bx}
B.~I. Zwiebel, ``{From Scattering Amplitudes to the Dilatation Generator in
  $\cN=4$ SYM},'' \href{http://dx.doi.org/10.1088/1751-8113/45/11/115401}{{\em
  J.Phys.} {\bfseries A45} (2012) 115401},
\href{http://arxiv.org/abs/1111.0083}{{\ttfamily arXiv:1111.0083 [hep-th]}}.

\bibitem{Bern:1994zx}
Z.~Bern, L.~J. Dixon, D.~C. Dunbar, and D.~A. Kosower, ``{One-loop $n$-point
  gauge theory amplitudes, unitarity and collinear limits},''
  \href{http://dx.doi.org/10.1016/0550-3213(94)90179-1}{{\em Nucl.Phys.}
  {\bfseries B425} (1994) 217--260},
\href{http://arxiv.org/abs/hep-ph/9403226}{{\ttfamily arXiv:hep-ph/9403226
  [hep-ph]}}.

\bibitem{Bern:1994cg}
Z.~Bern, L.~J. Dixon, D.~C. Dunbar, and D.~A. Kosower, ``{Fusing gauge theory
  tree amplitudes into loop amplitudes},''
  \href{http://dx.doi.org/10.1016/0550-3213(94)00488-Z}{{\em Nucl.Phys.}
  {\bfseries B435} (1995) 59--101},
\href{http://arxiv.org/abs/hep-ph/9409265}{{\ttfamily arXiv:hep-ph/9409265
  [hep-ph]}}.

\bibitem{Britto:2004nc}
R.~Britto, F.~Cachazo, and B.~Feng, ``{Generalized unitarity and one-loop
  amplitudes in $\cN=4$ super-Yang-Mills},''
  \href{http://dx.doi.org/10.1016/j.nuclphysb.2005.07.014}{{\em Nucl.Phys.}
  {\bfseries B725} (2005) 275--305},
\href{http://arxiv.org/abs/hep-th/0412103}{{\ttfamily arXiv:hep-th/0412103
  [hep-th]}}.

\bibitem{Brandhuber:2015boa}
A.~Brandhuber, B.~Penante, G.~Travaglini, and D.~Young, ``{Integrability and
  unitarity},'' \href{http://dx.doi.org/10.1007/JHEP05(2015)005}{{\em JHEP}
  {\bfseries 05} (2015) 005},
\href{http://arxiv.org/abs/1502.06627}{{\ttfamily arXiv:1502.06627 [hep-th]}}.

\bibitem{Engelund:2015cfa}
O.~T. Engelund, ``{Lagrangian Insertion in the Light-Like Limit and the
  Super-Correlators/Super-Amplitudes Duality},''
\href{http://arxiv.org/abs/1502.01934}{{\ttfamily arXiv:1502.01934 [hep-th]}}.

\bibitem{Koster:2014fva}
L.~Koster, V.~Mitev, and M.~Staudacher, ``{A Twistorial Approach to
  Integrability in $\mathcal{N}=4$ SYM},''
  \href{http://dx.doi.org/10.1002/prop.201400085}{{\em Fortsch.Phys.}
  {\bfseries 63} no.~2, (2015) 142--147},
\href{http://arxiv.org/abs/1410.6310}{{\ttfamily arXiv:1410.6310 [hep-th]}}.

\bibitem{Chicherin:2014uca}
D.~Chicherin, R.~Doobary, B.~Eden, P.~Heslop, G.~P. Korchemsky, L.~Mason, and
  E.~Sokatchev, ``{Correlation functions of the chiral stress-tensor multiplet
  in $ \mathcal{N}=4 $ SYM},''
  \href{http://dx.doi.org/10.1007/JHEP06(2015)198}{{\em JHEP} {\bfseries 06}
  (2015) 198},
\href{http://arxiv.org/abs/1412.8718}{{\ttfamily arXiv:1412.8718 [hep-th]}}.

\bibitem{Brandhuber:2014pta}
A.~Brandhuber, B.~Penante, G.~Travaglini, and D.~Young, ``{Integrability and
  MHV diagrams in $\mathcal{N}=4$ supersymmetric Yang-Mills theory},''
  \href{http://dx.doi.org/10.1103/PhysRevLett.114.071602}{{\em Phys.Rev.Lett.}
  {\bfseries 114} (2015) 071602},
\href{http://arxiv.org/abs/1412.1019}{{\ttfamily arXiv:1412.1019 [hep-th]}}.

\bibitem{Beisert:2010jr}
N.~Beisert {\em et~al.}, ``{Review of AdS/CFT Integrability: An Overview},''
  \href{http://dx.doi.org/10.1007/s11005-011-0529-2}{{\em Lett.Math.Phys.}
  {\bfseries 99} (2012) 3--32},
\href{http://arxiv.org/abs/1012.3982}{{\ttfamily arXiv:1012.3982 [hep-th]}}.

\bibitem{Minahan:2002ve}
J.~Minahan and K.~Zarembo, ``{The Bethe ansatz for $\cN=4$ superYang-Mills},''
  {\em JHEP} {\bfseries 0303} (2003) 013,
\href{http://arxiv.org/abs/hep-th/0212208}{{\ttfamily arXiv:hep-th/0212208
  [hep-th]}}.

\bibitem{Beisert:2003tq}
N.~Beisert, C.~Kristjansen, and M.~Staudacher, ``{The Dilatation operator of
  conformal $\mathcal{N}=4$ superYang-Mills theory},''
  \href{http://dx.doi.org/10.1016/S0550-3213(03)00406-1}{{\em Nucl.Phys.}
  {\bfseries B664} (2003) 131--184},
\href{http://arxiv.org/abs/hep-th/0303060}{{\ttfamily arXiv:hep-th/0303060
  [hep-th]}}.

\bibitem{Sieg:2010tz}
C.~Sieg, ``{Superspace computation of the three-loop dilatation operator of
  $\cN=4$ SYM theory},''
  \href{http://dx.doi.org/10.1103/PhysRevD.84.045014}{{\em Phys.Rev.}
  {\bfseries D84} (2011) 045014},
\href{http://arxiv.org/abs/1008.3351}{{\ttfamily arXiv:1008.3351 [hep-th]}}.

\bibitem{Bargheer:2008jt}
T.~Bargheer, N.~Beisert, and F.~Loebbert, ``{Boosting Nearest-Neighbour to
  Long-Range Integrable Spin Chains},''
  \href{http://dx.doi.org/10.1088/1742-5468/2008/11/L11001}{{\em J.Stat.Mech.}
  {\bfseries 0811} (2008) L11001},
\href{http://arxiv.org/abs/0807.5081}{{\ttfamily arXiv:0807.5081 [hep-th]}}.

\bibitem{Bargheer:2009xy}
T.~Bargheer, N.~Beisert, and F.~Loebbert, ``{Long-Range Deformations for
  Integrable Spin Chains},''
  \href{http://dx.doi.org/10.1088/1751-8113/42/28/285205}{{\em J.Phys.}
  {\bfseries A42} (2009) 285205},
\href{http://arxiv.org/abs/0902.0956}{{\ttfamily arXiv:0902.0956 [hep-th]}}.

\bibitem{Mueller:1979ih}
A.~H. Mueller, ``{On the Asymptotic Behavior of the Sudakov Form-factor},''
\href{http://dx.doi.org/10.1103/PhysRevD.20.2037}{{\em Phys.Rev.} {\bfseries
  D20} (1979) 2037}.

\bibitem{Collins:1980ih}
J.~C. Collins, ``{Algorithm to Compute Corrections to the Sudakov
  Form-factor},''
\href{http://dx.doi.org/10.1103/PhysRevD.22.1478}{{\em Phys.Rev.} {\bfseries
  D22} (1980) 1478}.

\bibitem{Sen:1981sd}
A.~Sen, ``{Asymptotic Behavior of the Sudakov Form-Factor in QCD},''
\href{http://dx.doi.org/10.1103/PhysRevD.24.3281}{{\em Phys.Rev.} {\bfseries
  D24} (1981) 3281}.

\bibitem{Magnea:1990zb}
L.~Magnea and G.~F. Sterman, ``{Analytic continuation of the Sudakov
  form-factor in QCD},''
\href{http://dx.doi.org/10.1103/PhysRevD.42.4222}{{\em Phys.Rev.} {\bfseries
  D42} (1990) 4222--4227}.

\bibitem{Anastasiou:2003kj}
C.~Anastasiou, Z.~Bern, L.~J. Dixon, and D.~Kosower, ``{Planar amplitudes in
  maximally supersymmetric Yang-Mills theory},''
  \href{http://dx.doi.org/10.1103/PhysRevLett.91.251602}{{\em Phys.Rev.Lett.}
  {\bfseries 91} (2003) 251602},
\href{http://arxiv.org/abs/hep-th/0309040}{{\ttfamily arXiv:hep-th/0309040
  [hep-th]}}.

\bibitem{Bern:2005iz}
Z.~Bern, L.~J. Dixon, and V.~A. Smirnov, ``{Iteration of planar amplitudes in
  maximally supersymmetric Yang-Mills theory at three loops and beyond},''
  \href{http://dx.doi.org/10.1103/PhysRevD.72.085001}{{\em Phys.Rev.}
  {\bfseries D72} (2005) 085001},
\href{http://arxiv.org/abs/hep-th/0505205}{{\ttfamily arXiv:hep-th/0505205
  [hep-th]}}.

\bibitem{Catani:1998bh}
S.~Catani, ``{The Singular behavior of QCD amplitudes at two loop order},''
  \href{http://dx.doi.org/10.1016/S0370-2693(98)00332-3}{{\em Phys.Lett.}
  {\bfseries B427} (1998) 161--171},
\href{http://arxiv.org/abs/hep-ph/9802439}{{\ttfamily arXiv:hep-ph/9802439
  [hep-ph]}}.

\bibitem{Sterman:2002qn}
G.~F. Sterman and M.~E. Tejeda-Yeomans, ``{Multiloop amplitudes and
  resummation},'' \href{http://dx.doi.org/10.1016/S0370-2693(02)03100-3}{{\em
  Phys.Lett.} {\bfseries B552} (2003) 48--56},
\href{http://arxiv.org/abs/hep-ph/0210130}{{\ttfamily arXiv:hep-ph/0210130
  [hep-ph]}}.

\bibitem{Bartels:2008ce}
J.~Bartels, L.~Lipatov, and A.~Sabio~Vera, ``{BFKL Pomeron, Reggeized gluons
  and Bern-Dixon-Smirnov amplitudes},''
  \href{http://dx.doi.org/10.1103/PhysRevD.80.045002}{{\em Phys.Rev.}
  {\bfseries D80} (2009) 045002},
\href{http://arxiv.org/abs/0802.2065}{{\ttfamily arXiv:0802.2065 [hep-th]}}.

\bibitem{Bern:2008ap}
Z.~Bern, L.~Dixon, D.~Kosower, R.~Roiban, M.~Spradlin, C.~Vergu, and
  A.~Volovich, ``{The Two-Loop Six-Gluon MHV Amplitude in Maximally
  Supersymmetric Yang-Mills Theory},''
  \href{http://dx.doi.org/10.1103/PhysRevD.78.045007}{{\em Phys.Rev.}
  {\bfseries D78} (2008) 045007},
\href{http://arxiv.org/abs/0803.1465}{{\ttfamily arXiv:0803.1465 [hep-th]}}.

\bibitem{Drummond:2008aq}
J.~Drummond, J.~Henn, G.~Korchemsky, and E.~Sokatchev, ``{Hexagon Wilson loop =
  six-gluon MHV amplitude},''
  \href{http://dx.doi.org/10.1016/j.nuclphysb.2009.02.015}{{\em Nucl.Phys.}
  {\bfseries B815} (2009) 142--173},
\href{http://arxiv.org/abs/0803.1466}{{\ttfamily arXiv:0803.1466 [hep-th]}}.

\bibitem{CaronHuot:2012ab}
S.~Caron-Huot and K.~J. Larsen, ``{Uniqueness of two-loop master contours},''
  \href{http://dx.doi.org/10.1007/JHEP10(2012)026}{{\em JHEP} {\bfseries 1210}
  (2012) 026},
\href{http://arxiv.org/abs/1205.0801}{{\ttfamily arXiv:1205.0801 [hep-ph]}}.

\bibitem{ArkaniHamed:2012nw}
N.~Arkani-Hamed, J.~L. Bourjaily, F.~Cachazo, A.~B. Goncharov, A.~Postnikov,
  and J.~Trnka, ``{Scattering Amplitudes and the Positive Grassmannian},''
\href{http://arxiv.org/abs/1212.5605}{{\ttfamily arXiv:1212.5605 [hep-th]}}.

\bibitem{Nandan:2013ip}
D.~Nandan, M.~F. Paulos, M.~Spradlin, and A.~Volovich, ``{Star Integrals,
  Convolutions and Simplices},''
  \href{http://dx.doi.org/10.1007/JHEP05(2013)105}{{\em JHEP} {\bfseries 1305}
  (2013) 105},
\href{http://arxiv.org/abs/1301.2500}{{\ttfamily arXiv:1301.2500 [hep-th]}}.

\bibitem{Kotikov:2001sc}
A.~Kotikov and L.~Lipatov, ``{DGLAP and BFKL evolution equations in the
  $\mathcal{N}=4$ supersymmetric gauge theory},''
\href{http://arxiv.org/abs/hep-ph/0112346}{{\ttfamily arXiv:hep-ph/0112346
  [hep-ph]}}.

\bibitem{Kotikov:2004er}
A.~Kotikov, L.~Lipatov, A.~Onishchenko, and V.~Velizhanin, ``{Three loop
  universal anomalous dimension of the Wilson operators in $\mathcal{N}=4$ SUSY
  Yang-Mills model},''
  \href{http://dx.doi.org/10.1016/j.physletb.2004.05.078}{{\em Phys.Lett.}
  {\bfseries B595} (2004) 521--529},
\href{http://arxiv.org/abs/hep-th/0404092}{{\ttfamily arXiv:hep-th/0404092
  [hep-th]}}.

\bibitem{Kotikov:2006ts}
A.~Kotikov and L.~Lipatov, ``{On the highest transcendentality in
  $\mathcal{N}=4$ SUSY},''
  \href{http://dx.doi.org/10.1016/j.nuclphysb.2007.01.020}{{\em Nucl.Phys.}
  {\bfseries B769} (2007) 217--255},
\href{http://arxiv.org/abs/hep-th/0611204}{{\ttfamily arXiv:hep-th/0611204
  [hep-th]}}.

\bibitem{Li:2014afw}
Y.~Li, A.~von Manteuffel, R.~M. Schabinger, and H.~X. Zhu, ``{Soft-virtual
  corrections to Higgs production at N$^3$LO},''
  \href{http://dx.doi.org/10.1103/PhysRevD.91.036008}{{\em Phys.Rev.}
  {\bfseries D91} no.~3, (2015) 036008},
\href{http://arxiv.org/abs/1412.2771}{{\ttfamily arXiv:1412.2771 [hep-ph]}}.

\bibitem{Gehrmann:2011aa}
T.~Gehrmann, M.~Jaquier, E.~Glover, and A.~Koukoutsakis, ``{Two-Loop QCD
  Corrections to the Helicity Amplitudes for $H \to$ 3 partons},''
  \href{http://dx.doi.org/10.1007/JHEP02(2012)056}{{\em JHEP} {\bfseries 1202}
  (2012) 056},
\href{http://arxiv.org/abs/1112.3554}{{\ttfamily arXiv:1112.3554 [hep-ph]}}.

\bibitem{Belitsky:2013ofa}
A.~Belitsky, S.~Hohenegger, G.~Korchemsky, E.~Sokatchev, and A.~Zhiboedov,
  ``{Energy-Energy Correlations in $\mathcal{N}=4$ Supersymmetric Yang-Mills
  Theory},'' \href{http://dx.doi.org/10.1103/PhysRevLett.112.071601}{{\em
  Phys.Rev.Lett.} {\bfseries 112} no.~7, (2014) 071601},
\href{http://arxiv.org/abs/1311.6800}{{\ttfamily arXiv:1311.6800 [hep-th]}}.

\bibitem{Nair:1988bq}
V.~Nair, ``{A Current Algebra for Some Gauge Theory Amplitudes},''
\href{http://dx.doi.org/10.1016/0370-2693(88)91471-2}{{\em Phys.Lett.}
  {\bfseries B214} (1988) 215}.

\bibitem{Bargheer:2009qu}
T.~Bargheer, N.~Beisert, W.~Galleas, F.~Loebbert, and T.~McLoughlin,
  ``{Exacting $\cN=4$ Superconformal Symmetry},''
  \href{http://dx.doi.org/10.1088/1126-6708/2009/11/056}{{\em JHEP} {\bfseries
  0911} (2009) 056},
\href{http://arxiv.org/abs/0905.3738}{{\ttfamily arXiv:0905.3738 [hep-th]}}.

\bibitem{Smirnov:2004ym}
V.~A. Smirnov, ``{Evaluating Feynman integrals},''
{\em Springer Tracts Mod.Phys.} {\bfseries 211} (2004) 1--244.

\bibitem{Lee:2013mka}
R.~N. Lee, ``{LiteRed 1.4: a powerful tool for reduction of multiloop
  integrals},'' \href{http://dx.doi.org/10.1088/1742-6596/523/1/012059}{{\em
  J.Phys.Conf.Ser.} {\bfseries 523} (2014) 012059},
\href{http://arxiv.org/abs/1310.1145}{{\ttfamily arXiv:1310.1145 [hep-ph]}}.

\bibitem{Gehrmann:2000zt}
T.~Gehrmann and E.~Remiddi, ``{Two loop master integrals for $\gamma^*$
  $\longrightarrow$ 3 jets: The Planar topologies},''
  \href{http://dx.doi.org/10.1016/S0550-3213(01)00057-8}{{\em Nucl.Phys.}
  {\bfseries B601} (2001) 248--286},
\href{http://arxiv.org/abs/hep-ph/0008287}{{\ttfamily arXiv:hep-ph/0008287
  [hep-ph]}}.

\bibitem{Goncharov09}
A.~B. Goncharov, ``{A simple construction of Grassmannian polylogarithms},''
\href{http://arxiv.org/abs/0908.2238}{{\ttfamily arXiv:0908.2238 [math.AG]}}.

\bibitem{Goncharov:2010jf}
A.~B. Goncharov, M.~Spradlin, C.~Vergu, and A.~Volovich, ``{Classical
  Polylogarithms for Amplitudes and Wilson Loops},''
  \href{http://dx.doi.org/10.1103/PhysRevLett.105.151605}{{\em Phys.Rev.Lett.}
  {\bfseries 105} (2010) 151605},
\href{http://arxiv.org/abs/1006.5703}{{\ttfamily arXiv:1006.5703 [hep-th]}}.

\bibitem{VerguCode}
C.~Vergu, ``Lecture Notes for the Mathematica Summer School on Theoretical
  Physics.'' \url{http://msstp.org/sites/default/files/Demo.nb}, 2011.

\bibitem{Bern:1991aq}
Z.~Bern and D.~A. Kosower, ``{The Computation of loop amplitudes in gauge
  theories},''
\href{http://dx.doi.org/10.1016/0550-3213(92)90134-W}{{\em Nucl.Phys.}
  {\bfseries B379} (1992) 451--561}.

\bibitem{Beisert:2003ys}
N.~Beisert, ``{The su(2$|$3) dynamic spin chain},''
  \href{http://dx.doi.org/10.1016/j.nuclphysb.2003.12.032}{{\em Nucl.Phys.}
  {\bfseries B682} (2004) 487--520},
\href{http://arxiv.org/abs/hep-th/0310252}{{\ttfamily arXiv:hep-th/0310252
  [hep-th]}}.

\bibitem{Beisert:2010gn}
N.~Beisert, J.~Henn, T.~McLoughlin, and J.~Plefka, ``{One-Loop Superconformal
  and Yangian Symmetries of Scattering Amplitudes in $\mathcal{N}=4$ Super
  Yang-Mills},'' \href{http://dx.doi.org/10.1007/JHEP04(2010)085}{{\em JHEP}
  {\bfseries 1004} (2010) 085},
\href{http://arxiv.org/abs/1002.1733}{{\ttfamily arXiv:1002.1733 [hep-th]}}.

\bibitem{Grisaru:1976vm}
M.~T. Grisaru, H.~Pendleton, and P.~van Nieuwenhuizen, ``{Supergravity and the
  S Matrix},''
\href{http://dx.doi.org/10.1103/PhysRevD.15.996}{{\em Phys.Rev.} {\bfseries
  D15} (1977) 996}.

\end{thebibliography}\endgroup

\end{fmffile}
\end{document}